\documentclass{article}
\usepackage{fullpage}
\usepackage{amsmath,amsthm,amssymb}
\usepackage{natbib}
\usepackage{graphicx}
\usepackage{hyperref}

\renewcommand{\P}{\mathbb{P}}
\newcommand{\E}{\mathbb{E}}
\newcommand{\R}{\mathbb{R}}

\begin{document}

\title{Parallel adaptation: One or many waves of advance of an advantageous allele?}
\author{Peter Ralph$^{1}$,Graham Coop$^{1}$\\
\small $^1$ Department of Evolution and Ecology \& Center for Population Biology,\\
\small University of California, Davis.\\
\small To whom correspondence should be addressed: \texttt{plralph@ucdavis.edu}\\
}
\date{}

\maketitle

\section*{Abstract}
Models for detecting the effect of adaptation on population genomic diversity
are often predicated on a single newly arisen mutation sweeping rapidly to fixation. 
However, a population can also adapt to a new environment by multiple
mutations of similar phenotypic effect that arise in parallel, at the same locus
or different loci. 
These mutations can each quickly reach intermediate frequency, preventing any single one from rapidly sweeping to fixation globally, leading to a ``soft'' sweep in the population. 
Here we study various models of parallel mutation in a continuous, geographically spread population adapting to a global
selection pressure. 
The slow geographic spread of a selected allele due to limited dispersal 
can allow other selected alleles to arise and start to spread elsewhere in the species range. 
When these different selected alleles meet, their spread can slow dramatically, 
and so initially form a geographic patchwork, a random
tessellation, which could be mistaken for a signal of local adaptation. 
This spatial tessellation will dissipate over time due to mixing by migration, leaving a set of partial sweeps within the global population. 
We show that the spatial tessellation initially formed by mutational types is closely connected to Poisson
process models of crystallization, which we extend.  
We find that the probability of parallel mutation and the spatial scale on which parallel mutation occurs is captured by a single compound parameter, 
a characteristic length, that reflects the expected distance a spreading allele travels before it encounters a different spreading allele. 
This characteristic length depends on the mutation rate, the dispersal parameter, the effective local density of individuals, and to a much lesser extent the strength of selection. 
While our knowledge of these parameters is poor, we argue that even in widely dispersing species, such parallel geographic sweeps may be surprisingly common. 
Thus, we predict that as more data becomes available, many more examples of intra-species parallel adaptation will be uncovered.

\section{Introduction}

There are many dramatic examples of convergent evolution across distantly related species, where a phenotype independently evolves via parallel changes at orthologous genetic loci \citep{Wood:2005,Arendt:08}, indicating that adaptation can be strongly shaped by pleiotropic constraints \citep{Haldane:book,Orr:05,Stern:08, Kopp:09}. There are also a growing number of examples of the parallel evolution of a phenotype {\it within} a species due to independent mutations at the same gene \citep{Arendt:08} (which are sometimes referred to as genetically redundant). Some of the best-studied examples come from the repeated evolution of resistance to insecticides within several insect species \citep{cyclodiene-resistance}, and the resistance of malaria to antimalarial drugs \citep{Anderson:05,pearce2009malaria}. Another example is the loss of pigmentation in {\it Drosophlia santomea} through least three independent mutations at a {\it cis}-regulatory element \citep{jeong-pigmentation}, while the evolution of pigmentation within vertebrate species provides further examples \citep{peromyscus-pigmentation, gross-cavefish-albinism, protas-cavefish-albinism}. There are also a number of examples of parallel evolution within our own species \citep{Novembre:09}. For example, various G6PD mutations have spread in parallel in response to malaria \citep{tishkoff-g6pd, G6PD:09}, and lactase-persistence has evolved independently in at least three different pastoral populations \citep{enattah-milk-culture,tishkoff-lactase}. A particularly impressive example in humans is offered by the sickle cell allele at the $\beta$-globin gene that confers malaria resistance, where multiple changes have putatively occurred at a single base pair \citep[see][for discussion]{Flint:98}. 
In each of these examples, multiple, independent mutations have lead to the same or functionally equivalent adaptive phenotype, although they differ in the degree to which the functional consequences and equivalences of the different mutations have been explored. Such repeated adaptive evolution via similar changes within a species, which we term {\em parallel adaptation}, may therefore be common. As we will also address repeated evolution of a similar phenotype via changes at different genetic loci this could more broadly be termed ``convergent adaptation'' \citep{Arendt:08}.  

In many of these examples the selection pressure is patchy and rates of gene flow are low, increasing the chance of parallel adaptation. However, parallel adaptation can occur even in a panmictic population. For example, adaptation may occur from multiple independent copies of the selected allele present in standing variation at mutation-selection balance within the population \citep{Orr:01,softsweepsI}. Even when there is no standing variation for a trait in a panmictic population, a selected allele could arise independently several times during the course of a selective sweep, if mutation is sufficiently fast relative to the spread of the selected allele. This idea was formalized by \citet{softsweepsII, softsweepsIII}, who showed that such soft sweeps may be expected when the population scaled mutation rate (the product of the effective population size and mutation rate) towards the adaptive allele is $>1$. Thus, repeated mutation may be quite common for species with large populations, or where the mutation target is large, e.g. knocking out of a gene. Pennings and Hermisson showed that the number of independently arisen selected alleles in a sample has approximately the Ewens distribution \citep{softsweepsII}, and properties of neutral variation at a closely linked site can be derived from this \citep{softsweepsIII}. Such a selective sweep has been termed a {\em soft sweep}, as the population can adapt without the dramatic reduction in diversity at linked selected sites that is usually associated with a full sweep \citep{MaynardSmithAndHaigh74}, see \citet{Hermisson2008}, \citet{softsweepsII}, \citet{softsweepsIII}, and \citet{Pritchard:10} for discussion, and \citet{Schlenke:05} or \citet{jeong-pigmentation} for potential examples.
 
Clearly, if parallel mutations can occur during adaptation in a large panmictic population, then limited dispersal should further increase the chance of parallel adaptation, as other mutations can arise and spread during the time it takes one to move across the species range. 
Intuitively, a low rate of dispersal and a large mutational target should increase the chance of parallel adaptation \citep[as in][]{Coop:09,Novembre:09}, 
but it is unclear exactly how other dispersal, population and mutational parameters play into the probability of parallel adaptation. 
However, in the absence of a formal model, many simple questions remain: Does parallel adaptation only occur in species with strong population structure? 
Weak selection pressures lead to slowly spreading mutations; is parallel adaptation more likely in this case?
This leaves us unable to understand the likelihood of parallel adaptation in particular examples \citep[such as][]{Flint:98} and more generally its role in geographic patterns of adaptation \citep[such as][]{Coop:09}.

Here we study parallel adaptation in a homogeneous, geographically spread population. We focus on the case where a population is exposed to a novel selection regime throughout a homogeneous species range, and the population is initially entirely devoid of standing variation for the trait, assumptions that favor the fixation of only a single new allele in the population. We use simple approximations to derive theoretical results for the properties of parallel adaptation in a continuous spatial population with strong migration for a range of dispersal distributions (also called dispersal kernels, including fat-tailed examples).  We are able to describe fairly completely the resulting patterns, and show that they are well captured by a single compound parameter combining the rate of mutation and the speed at which the mutation spreads. For an introduction to the patterns of genetic diversity that can be expected from such geographic structure at both neutral and selected loci, see \citet{Charlesworth:03}, \citet{Novembre:09}, and \citet{Lenormand:02}. 

We show that when population sizes are sufficiently large and dispersal distances are small compared to the species range, parallel adaptation within a species is likely to be common, and quantify this relationship. Furthermore, we describe how separately-arisen mutations will--- at least for some time--- leave behind a spatial pattern reflecting their separate origins. 

The structure of this paper is as follows. In Section \ref{ss:continuous} we introduce and analyze our model of a continuous population,
first in the classical context, and then in a more general context that allows for accelerating waves (arising from fat-tailed dispersal distributions).
In Section \ref{ss:simulations} we present the results of some simulations of the continuous process, intended to assess the robustness of our results to deviations from the assumptions.
In Sections \ref{ss:continuousresults} and \ref{ss:numerics} we present and discuss the theoretical results in a few biologically reasonable contexts, providing numerical results to illustrate how the different parameters play into the probability of parallel adaptation. 
In Section \ref{ss:discussion} we discuss consequences and extensions.
Some mathematical arguments are postponed until the Appendix.


\subsection{Modeling assumptions}
Here we describe the assumptions behind our model and give some background, before introducing in Section \ref{ss:continuous} the model we analyze.
First, we assume each mutation under consideration confers a selective advantage such that,
upon appearing in the population,
it quickly rises locally to some equilibrium frequency.
Second, there is significant spatial structure,
namely, migration is weak enough that the selected trait reaches an equilibrium frequency locally
before spreading to the entire population.
Third, the parallel mutations are distinguishable, and confer the same selective benefit.
Fourth, these mutations are neutral relative to each other, 
in the sense that in a population at equilibrium frequency (e.g. fixation) for any collection of these mutations,
the dynamics of their relative proportions occur on a longer time scale than their dynamics in the original background (examples are given below).
We call this last assumption {\em allelic exclusion}, since it implies that areas fixed for one adaptive allele will not be rapidly overtaken by another.

Under these assumptions, a newly arisen advantageous mutation, if it is initially successful,
will spread through the population in a more-or-less wavelike manner (more on this later).
If another allele conferring the same advantage arises in a location the first has not yet reached,
then the two waves spread towards each other and will at some point collide.
What happens when they collide will generally depend on the details of their epistatic interaction, 
or, if they occur at a single site, on their dominance interaction. 
However, by our assumption of allelic exclusion,
the dynamics are slower than the spread of the selected alleles.
This allows us to neglect the slower mixing of types and genetic drift
that will happen in this phase, instead focusing on the first process
by which independently arisen alleles partition the population.

\begin{figure}[ht]
\begin{center}
\includegraphics[width=1.0\textwidth]{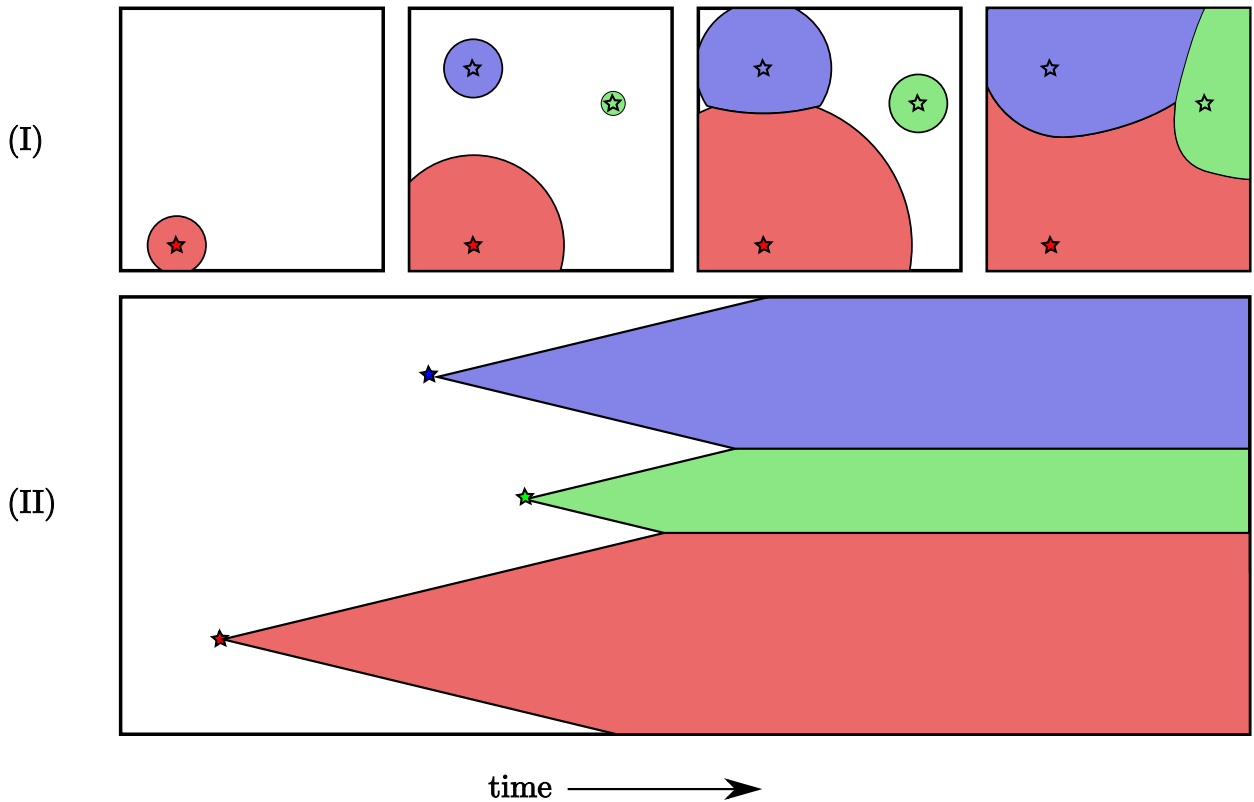}
\caption{ %
{\bf A cartoon representation of our model of spatial parallel mutation.} In the top row, each panel represents a 2D species range with time increasing from the left to right panel. In the bottom row, a 1D species range is represented by the vertical axis and time is the horizontal axis, with more recent times closer to the right side of the page. Stars represent a new mutation arising and escaping drift. The three colors represent the area occupied by three different alleles.  Note that (I) and (II) are not different views of the same process, although they are similar.
}
\label{fig:cartoon}
\end{center}
\end{figure}

In Figure \ref{fig:cartoon} we show a cartoon to illustrate our model, and in Figures \ref{fig:1dsim2} and \ref{fig:2dsim} we show the results of a simulation (described in Section \ref{ss:simulations}).

\paragraph{Allelic exclusion}
The allelic exclusion assumption is fundamental to our approach. 
It will hold, for instance, if there is a single advantageous mutation, 
and we treat each time it arises independently as a distinct allele, identifiable by examination of linked neutral variation. 
It will also hold if mutations at multiple sites within a gene are genetically redundant, 
such as loss of function mutations, 
and no additional selective benefit is conferred by having a mutation at more than one site 
(though this may be an approximation, since even loss of function changes within the same gene may differ in their characteristics, as in \citet{Rosenblum:10}).

Another important consequence of allelic exclusion
is that a mutation occurring in a location where the advantageous allele
already exists in large numbers is unlikely to persist or achieve high frequency--- 
indeed, if the interaction is neutral and there already exist in the same location 999 other individuals with the selected trait,
then a new mutation will contribute on average only .001 of the future population,
and has high probability of being lost from the population by drift.
This fact allows us to ignore all new mutations that occur after any selected allele
has risen in that location to a nonzero frequency.
In particular, the shape of the wave front will not be important,
only how its leading edge spreads. 
Below, for convenience we often talk about the probability or rate of local {\em fixation}, but it follows from this observation that we need only require that the allele escapes loss from the population by drift and that some intermediate equilibrium frequency is reached, as would occur in the case of overdominance.

\paragraph{Selection}
We also assume that the advantageous, derived alleles have a reproductive advantage of $(1+s)$ relative to the ancestral type.
In practice, in a diploid model with dominance or epistasis, or in the presence of density dependence, 
we require that both the manner in which a new mutation escapes drift,
and the way that it subsequently spreads through the population,
be well-approximated by the simple haploid (or additive) model.
Roughly speaking, this holds if the growth and spread of the allele is driven by growth 
where the allele is at very low frequency (and primarily occurring in heterozygotes).
This implies that the probability a new mutation escapes drift is well-approximated by $2s$ divided by the variance in offspring number
(which is quite robust to the details of spatial structure \citep{Maruyama:70,Maruyama:74}),
and that per-capita growth is fastest when at low frequency.
In the usual formulation of diploid systems \citep{aronson1978multidimensional}, 
this is satisfied if the fitness advantage of the homozygote is no more than twice the fitness advantage of the heterozygote.
In other cases, e.g.\ an Allee effect, the behavior can be quite different; see \citet{stokes1976types}.

\subsection{Background on the wave of advance}

We model the spread of a selected allele by making use of existing work on traveling waves, 
a link first established independently by \citet{fisher1937wave} and by \citet*{KPP1937}.
We introduce and review the wave of advance literature here, as much of the subsequent development
has occurred in fields other than population genetics.
Suppose that individuals produce a random number of offspring with mean $r$, 
that offspring disperse a random distance with standard deviation $\sigma$,
and let $p(t,x)$ be the expected proportion of mutants at time $t$ and location $x$.
Suppose also that the selection coefficient $s$ is small and the advantage is additive, and that the population density $\rho$ is fairly large.
Both papers argued that if the dispersal distance is Gaussian, or if $\sigma$ is small (so that the ``long-time'' dispersal distribution is Gaussian),
then barring the appearance of new mutations, 
the time evolution of $p$ is well-described by the reaction--diffusion equation now known as the Fisher--KPP equation,
\begin{equation} \label{eqn:FKPP}
  \frac{\partial}{\partial t} p(t,x) 
    = \frac12 \left( r \sigma^2 \right) \sum_{i=1}^d \frac{\partial^2}{\partial x_i^2} p(t,x) 
        + \left( r s \right) p(t,x) ( 1-p(t,x) ),
\end{equation}
where $d$ is the dimension of the species range.
They furthermore showed in $d=1$ that a ``wave of advance'' occurs as the solution to this equation, 
and that for initial conditions where the allele is only polymorphic within a spatially bounded region,
the solution moved asymptotically with speed $v=r\sigma\sqrt{2s}$.
\citet{KPP1937} also covered the more general case in which $p(x,t)(1-p(x,t))$ is replaced by $F(p(x,t))$ for an appropriate function $F$, 
which gives the density dependent growth rate of the selected type, subject to certain conditions.

For many other choices of dispersal distribution and growth function $F$
the advancing front of a new type also approaches a constant wave shape that advances at constant speed through time--- a ``traveling wave'' solution,
but with a speed not given by the same formula.
Then the frequency of individuals of the selected type at $x$ at time $t$ can be expressed $p(x,t) = h(x-vt)$, 
where $h(\cdot)$ gives the shape of the wave and $v$ is its speed. 
These traveling wave solutions have been studied for the Fisher--KPP equation for a range of appropriate $F$ \citep{aronson1975nonlinear};
the speed can often be found more easily than the wave shape \citep{hadelerroethe}.
Radially symmetric solutions also exist, in which the new type travels outwards from an initial origin;
the behavior of such radially spreading waves depends on initial conditions,
but will asymptotically move with the same constant speed and fixed wave shape as in one dimension.

Since the introduction of the Fisher--KPP equation, traveling wave solutions to reaction--diffusion equations 
have been studied in the ecological literature as a model of invading species \citep{skellam1951dispersal,kot96}, as well as in a range of other fields. 
See \citet{aronson1978multidimensional} for some classical theory, general discussion and context, or \citet{volpert1994traveling} for a more extensive reference.
Related models, using integrodifference or integrodifferential equations 
have been used by various authors to include various important biological factors such as age structure and fluctuating environments
\citep{neubert2000invasion,neubert2000demography,kot2008saddlepoint}; see \citet{hastings2005spatial} or \citet{zhao2009spatial} for a review. 
Density regulation is often discussed in these models, but important behaviors can usually be determined by a linearization, based on how the new type grows when rare. 
Common to these models is the existence of traveling wave solutions, whose forms and speeds are often known only implicitly;
most natural models of the spread of a new selected type can be translated into one of these frameworks.
There is also a fruitful connection of these Fisher--KPP models to branching random walks that is beyond the scope of this article;
see \citet{mckean1975KPP,biggins1979growthrates,biggins1995growth} and \citet{Kot2004}.
A similar model, the contact process, has also been widely studied in the probabilistic literature;
see \citet{bramson1989crabgrass}.

The qualitative behavior of the spread of an organism or an allele in a population
can depend on the organism's dispersal kernel, defined as the probability density of the distance between mother and child's birth locations
(see \citet{shigesada1997invasions} or \citet{cousens2008dispersal} for discussion).
Most mathematical models of invasions assume that the dispersal kernel 
has tails bounded by an exponential, and obtain a constant wave speed.
In some species this is appropriate, while in others,
rare, long-distance migration events are important \citep{shigesada1997invasions}.
In such organisms, dispersal may be better modeled by a kernel that is not bounded by an exponential (i.e.\ a ``fat-tailed'' kernel),
although there is generally insufficient evidence so far \citep[][Ch.5]{cousens2008dispersal}.
\citet{mollison1972spatial} showed that in a certain model, if the kernel is fat-tailed 
the range occupied by the expanding type will be patchy and will grow faster than linearly: 
the spread accelerates and eventually moves faster than any constant-speed traveling wave. 
Moreover, \citet{pacalalewis:2000} have established a link between leptokurtic kernels 
(kernels whose kurtosis exceeds that of the standard Gaussian) and patchy invasion dynamics.
Leptokurtic but exponentially bounded kernels can lead to waves that initially accelerate but settle to a constant speed.
We shall see that the important behavior of the model is not determined by the asymptotic, long-time speed of the wave,
but rather its behavior at intermediate times.  
Therefore, kernels that have similar short-time behavior but different long-time behavior
can give rise to similar dynamics on the scale we are interested in.
Consideration of other wave behaviors leads to a more general model, which we study in Section \ref{ss:generalspeed}.

The models reviewed above are haploid models; traveling waves in diploid models have been much less studied.
\citet{aronson1975nonlinear} show that in the diploid analogue to Equation \eqref{eqn:FKPP},
if the difference in selection coefficient is small, then allele frequency dynamics are approximately governed by \eqref{eqn:FKPP}.
If local populations are in Hardy-Weinberg equilibrium, then more general results apply demonstrating the existence of traveling waves
\citep{weinberger1982longtime,zhao2009spatial}.
If dispersal occurs over a distance comparable to the width of the wave then this will no longer be the case,
and while recently developed general theory \citep{zhao2009spatial} might be applied,
the existence and characterization of traveling waves in other diploid models is to our knowledge an open question.
However, we certainly expect the behavior to be wavelike,
and since our theory takes wave behavior as an input,
we have no qualms about using our model to discuss the diploid organisms of Section \ref{ss:numerics}.

\section{Methods} \label{ss:continuous}

Consider a population with continuous spatial distribution. 
After the change initiating the novel selection regime (which occurs at $t=0$), selected mutations can arise at random throughout the species range and, 
if they escape loss due to genetic drift, start to expand radially out from their origination point. 
Under our assumption of mutational exclusion, a new mutation can only arise and spread in areas not already reached by another selected allele.

We first examine the case where the wave travels with constant velocity,
and then in Section \ref{ss:generalspeed} study the more general case where the speed of spread of the selected allele is not constant through time, 
proving our results in this more general setting.
It turns out that the properties of the final pattern of types can be conveniently summarized by a single compound parameter,
a {\em characteristic length}.
In the constant speed case, models with different parameters will result in patterns with identical properties
when the geographic distance is scaled in units of this characteristic length.
This is similar to the use of effective population size in models of genetic drift, 
for which populations with very different sizes have identical rates of drift when time is scaled by the effective population size.
This characteristic length can be defined as the distance traveled by an unobstructed spreading wave 
before it is expected that one other successful mutation would have arisen within the area so far enclosed.

Any such expanding waves of selected alleles in real populations are subject to stochastic fluctuations.  
Initially, we ignore this point, but in Section \ref{ss:stochasticwaves} we show that if stochasticity is taken into account,
first order properties such as the mean number of types only depend on the speed of the {\em mean} wave,
in a way we make explicit.

\subsection{The mutational process} \label{ss:mutationalprocess}

Imagine a large, haploid population with $\rho$ individuals per unit area distributed uniformly over some range $U$.  
The spatial range $U$ may be one- or two-dimensional, but must be connected (i.e.\ not composed of disjoint pieces).
We allow the number of offspring produced by each individual to be random, with mean $r$ and variance $\xi^2$,
and suppose that each offspring of a nonmutant carries some beneficial mutation with probability $\mu$.
Mutants have an additive advantage of $s$ relative to nonmutants and additional mutations have no effect;
to fix things, suppose that nonmutants reproduce at rate $1$ (so time is scaled in units of generations), while mutants reproduce at rate $1+s$,
and that each individual reproduces independently of the others and of the state of the population.
We will frequently make use of the fact that the probability of local fixation of a single new mutant,
since $s$ is assumed to be small, is well approximated by $2s/\xi^2$.

The set of times and locations $(t_i,x_i)$ at which new mutants appear is a random set of points in space-time.
Under our assumptions, it is well approximated by a Poisson point process in $[0,\infty) \times U$ 
with constant intensity $\rho \mu r$, the mutation rate per unit area per generation. 
(Recall that a Poisson point process with constant intensity is a random collection of points with the property that 
the number of points in any set has a Poisson distribution with mean equal to the area of the set multiplied by the intensity,
and the numbers of points in any two nonintersecting sets are independent of each other.)
By the ``thinning'' property of Poisson processes,
the points of origin of new mutations that will be successful if no other mutant type has already colonized the location
is also a Poisson point process in $[0,\infty)\times U$ with constant intensity $\lambda = (\rho \mu r) (2s / \xi^2)$ mutations per unit area per generation.

\subsection{Constant wave speed} \label{ss:constantspeed}

As reviewed earlier, if $\rho$ is large enough that stochastic effects are small,
the selected type will spread as an expanding wave with constant wave speed $v = r \sigma \sqrt{2s}$. 
The resulting picture is of waves spreading radially out from the site
of each successful mutation, until they collide with each other. 
Rather than formally prove the convergence of some discrete model, we will take this natural continuous model as our starting point for analysis:
successful mutations arise as the Poisson process described in \ref{ss:mutationalprocess}, and instantly begin to spread radially at speed $v$. 

This natural model of parallel adaptation has sources of new waves (new successful mutations) arising as a Poisson process in space and time,
and each wave expanding outwards until encountering another wave,
where they come to a halt (see Figure \ref{fig:cartoon}).
This model has been studied before--- again, first by \citet{kolmogorov-crystallization}---
as a model of crystallization,
in which nucleation sites form at random points in time and space 
and initiate the radial growth of new crystals.
More generally, the speed of the wave and the intensity of the Poisson process may not be constant,
in which case the final tessellation of space determined by the crystals is known as the
Kolmogorov-Johnson-Mehl-Avrami tessellation \citep{fanfoni-tomellini},
The properties of these crystal shapes have been extensively studied
in the case of constant-speed waves by \citet{moller92,moller95} and others \citep{bollobas-crystallization,gilbert-crystallization}.

To make this model precise, suppose that mutant type $i$ arose at location $x_i$ and time $t_i$.  At each later time $t$,
let $A_i(t)$ denote the area that type $i$ has spread to by time $t$ that was not reached first by any other type $j$.
This is formally defined by
\begin{equation} \label{eqn:Adefn}
    A_i(t) = \{ x \in U : \; \|x-x_i\| - v(t-t_i) \le \min\left(0,\|x-x_j\| - v(t-t_j)\right) \; \forall j \} ,
\end{equation}
where $\|x-y\|$ is the Euclidean distance from $x$ to $y$.
The dynamics of the collection of areas $A_i$, for origins $(t_i,x_i)$ distributed as a homogeneous Poisson process, is the (Kolmogorov-)Johnson-Mehl process.

One objection to this picture is that the waves do not really begin to spread instantly at speed $v$, 
they first need to reach equilibrium (e.g. fixation) locally,
which takes a random time of order $\log(\rho\sigma^d)/s$ generations,
and only converge to the equilibrium wave speed after the effect of initial conditions dies out.
Translating the points of a Poisson process by independent random amounts produces another Poisson process,
but this one will no longer be homogeneous in time. 
However, if the standard deviation of the time to local fixation is small relative to $\lambda$, this is has little effect.
Another objection is that the shape of the wave itself is stochastic, especially at the beginning.
We will ignore this detail and return to it in Section \ref{ss:stochasticwaves}. 
The simulations in Figure \ref{fig:takeoff}, and others for a range of parameters (not shown),
show that the waves quickly begin to spread at a constant speed, 
after a random delay with relatively low variance, reassuring us that our assumptions are reasonable.

In principle, any property of the model can now be found through calculations with Poisson processes,
although for general species ranges $U$ the formulas are complicated.
All results in this section can be derived more or less explicitly by giving a population genetics interpretation to results in \citet{moller92};
they will also follow from our results of Section \ref{ss:generalspeed}, where the proofs are given,
and which generalize the Johnson-Mehl model in a different direction than previously done by \citet{moller92}.

It turns out that all properties can be summarized fairly simply, especially in this constant speed case.
Recall that $\lambda = 2 s \rho \mu r / \xi^2$ is the spatial density of successful mutations per generation,
$v = \sigma \sqrt{2s}$ is the speed of the wave,
and let $\omega(d)$ be the area of the sphere of radius $1$ in $d$ dimensions, so $\omega(1)=2$ and $\omega(2)=\pi$.
We define the {\em characteristic length} $\chi$, which will be useful in a moment, as
\begin{equation} \label{eqn:constchidefn}
    \chi = \left( \frac{v}{\lambda \omega(d)} \right)^{1/(d+1)} = \left( \frac { \xi^2 \sigma } { \rho \mu \sqrt{2s} \omega(d) } \right)^{1/(d+1)} .
\end{equation}
Fix some habitat shape $U$ in $d$ dimensions ($d$ will be 1 or 2)
with diameter (the maximum distance between any two points in $U$) equal to $1$,
and let $U(a)$ be a habitat with the same shape, scaled by the factor $a$ (and hence diameter $a$).
Since the process with parameters $(v, \lambda, U(a))$ can be realized
by rescaling space by $1/a$ and time by $\lambda a^d$ in the process with parameters $(v/(a^{d+1} \lambda),1,U(1))$,
the distribution of the final configuration of types within $U(a)$
(specifically, $\{A_i(\infty)\}_i$)
is a function of the single number
\begin{equation}
    \frac{ \lambda a^{d+1} \omega(d) }{ v } = \frac{ \rho \mu \sqrt{2s} \omega(d) } { \xi^2 \sigma } a^{d+1} = \left( \frac{a}{\chi} \right)^{d+1},
\end{equation}
The quantity $(a/\chi)^{d+1}$ is, up to a constant,
the expected number of other mutations to arise in an area of diameter $a$ in the time it takes the wave to cover that area.
Furthermore, $\chi$ is the distance traveled by an unobstructed spreading wave before it is expected
that one other successful mutation would have arisen within the area enclosed so far.
(This last interpretation is the reason for the appearance of $\omega(d)$.)

A critical characteristic of parallel adaptation is the density of unique mutations per unit area.
This mean density of types $\nu(d)$ can be calculated exactly, and clearly displays the role of $\chi$.
We define $\nu(d)$ as the mean number of successful types arising in a region divided by the area of that region, 
assuming the species range is the infinite range $\R^d$ to avoid edge effects.
(This turns out to be independent of the region used.)
In two dimensions, we get from equation \eqref{eqn:generaldensity} (after the integrals worked out in Appendix \ref{apx:stretched}) that
\begin{equation}
    \nu(2) =  \chi^{-2} \frac{ \Gamma(4/3) 3^{1/3} }{ \pi } ,
\end{equation}
where $\Gamma$ is the gamma function,
and while in one dimension, $\nu(1) = \chi^{-1} 2^{-5/2}$.  
In general, the mean number of successful types in a $d$-dimensional region of total area $A$ will be
$A / \chi^d$, up to a constant depending only on the shape of the region.

The mean area occupied by a successful mutation in $d$ dimensions is $1/\nu(d)$.
In $d=2$ there does not seem to be a nice expression for the variance of this area,
but numerical computation and simulation indicate that the distribution is not highly skewed---
the areas occupied by different mutations are comparable (see e.g.\ Figures \ref{fig:1dsim2} and \ref{fig:2dsim}). 
\citet{moller92} also computes many other quantities in this constant-speed setting, 
such as the mean density of interfaces between adjacent types, or the mean number of neighbors of a given mutation--- we do not consider these here.

A related quantity is the distance from a sample point to the origin of the mutation that eventually covers it,
which relates to the total number of others sharing a sampled type.
The mean and variance (and all other moments) of this can be computed in a similar way to 
Equation \eqref{eqn:explmoments} in Appendix \ref{apx:stretched}.

Besides the spatial scale, the other most important datum is the time scale of adaptation.
Although the final distribution of types only depends on the characteristic length, 
the speed at which adaptation spreads through the population depends differently on the parameters.
A convenient summary of this time is $\tau$, defined as the time before a chosen point is hit by an expanding wave.
(For the point $x$, this is $\tau = \inf \{ t > 0 : x \in \bigcup_i A_i(t) \}$.)
In the infinite $d$--dimensional range $\R^d$, the time $\tau$ has a distribution such that $\tau^{d+1}$ is an exponential:
\[  
  \P \{ \tau > t \} = \exp \left( - \frac{ \lambda v^d \omega(d) }{ d+1 } t^{d+1} \right) ,
\]
so the expected time after selection begins until a chosen location has been reached by an adaptive mutation is
\begin{equation} \label{eqn:etau}
  \E[ \tau ] = \left( \lambda v^d \omega(d) \right)^{-1/(d+1)} \Gamma\left(\frac{1}{d+1}\right) (d+1)^{-d/(d+1)}
\end{equation}
Note that the crucial quantity here in, say, two dimensions, is
\[
    \left(\lambda v^2\right)^{-1/3} = \frac{ \xi^{2/3} }{ r (2s)^{2/3} (\rho \mu \sigma)^{1/3} } ,
\]
which depends much more strongly on the selection coefficient $s$ than does the characteristic length.


\subsection{The general case} \label{ss:generalspeed}

The results of the previous section can be applied to any model having waves of advance with a constant wave speed, substituting this speed for $v$.
However, real dispersing populations often show important effects of rare, long-distance dispersal events \citep[such as][]{coyne1982longdistance,clark1998trees}.
This implies that the dispersal distribution should in some sense be ``fat-tailed'', but since data on rare events is hard to obtain,
there is no consensus on what distributions best model real dispersal \citep{kot96}.
Since the behavior of the wave depends on the shape of the dispersal kernel--- waves from fatter-tailed kernels may tend to accelerate initially---
it is important to explore wave behaviors other than the simple constant-speed case \citep{kot96}.
This is more technically demanding, but we find that the spatial properties can again be meaningfully described by a single characteristic length.

As mentioned before, Mollison showed that if the dispersal kernel is not bounded by a decaying exponential (which we now refer to as ``fat-tailed''),
the range occupied by the expanding type expands at an ever-increasing speed \citep{mollison1972spatial}.  
Although the resulting behavior is termed an accelerating wave \citep{kot96}, 
there is no ``traveling wave solution'' in the same precise sense as before; 
thus, we imagine that the frequency $p(x,t)$ is radially symmetric and decreasing as $\|x\|$ increases for large enough $t$ and $\|x\|$;
and that for some fixed low frequency, $\epsilon$, the distance $f(t)$ at which the wave front first rises above a frequency $\epsilon$ 
(i.e.\ $f(t)=\sup \{ \|x\|>0 : p(x,t)>\epsilon \}$) is reasonably well-behaved.
In practice, we assume the existence of such an $f(t)$, and refer to this as the leading edge, and proceed from there. 
Note that it is not necessary to know the entire wave shape, only the speed of its leading edge, because of our assumption of mutation exclusion. 
Note also that $f$ depends on $\epsilon$, although in many cases the dependence only introduces a constant; see Section \ref{ss:numerics} for examples.

This leads to the following more general model.
As before, the population inhabits a region $U$ with uniform population density $\rho$,
and successful mutations occur at the points of a rate $\lambda = 2s\rho\mu r/\xi^2$ Poisson point process in $[0,\infty) \times U$ (time and space).
Any mutation that occurs in an unoccupied area begins to occupy the area around it, 
occupying by time $t$ any previously unoccupied points no more than distance $f(t)$ away,
where $f$ is an increasing function with $f(0)=0$ and $f(t) \to \infty$ as $t \to \infty$.
The function $f(t)$, the {\em wave expansion profile}, is the radius of an uninterrupted wave after time $t$, 
and a point at distance $r$ from the origin of a mutation will be reached after $f^{-1}(r)$ time units.

\begin{figure}[ht]
\begin{center}
\includegraphics[width=1.0\textwidth]{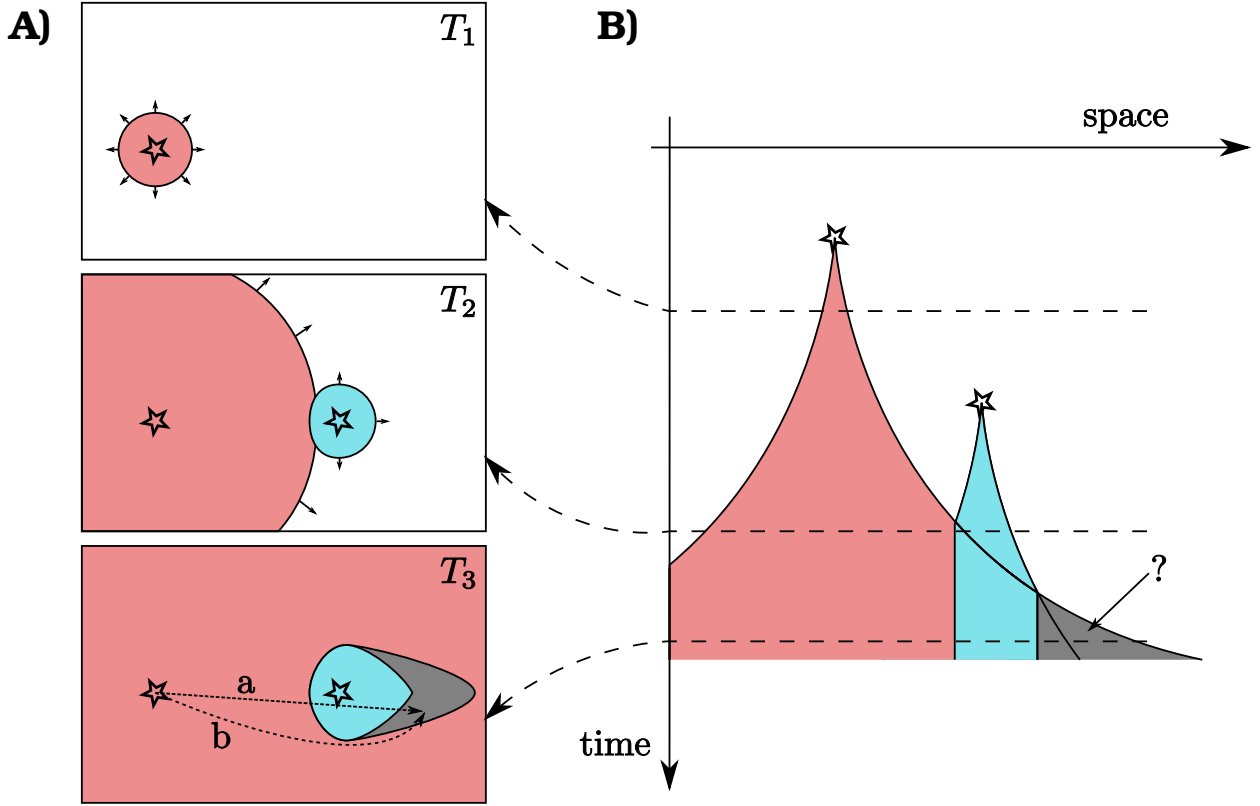}
\caption{ %
{\bf The delicate situation of accelerating waves.}  
{\bf A.} Three panels showing a 2D species range at three time points.
{\bf B.} A 1D cross section through the 2D range, with time along the vertical axis.
Stars represent new mutations, and colors code different types.
At time $T_1$, only the red mutation is present, and is expanding slowly.  
At time $T_2$, the red mutation is expanding quickly, but the blue mutation has just appeared and is still expanding slowly.
At time $T_3$, the red mutation has outflanked the blue mutation, but it is not clear which mutation should have claimed the grey area.
Each point in the grey area would have been reached first by the red mutation if unobstructed (trajectory $a$),
but if forced to detour around the blue mutation, the shortest path (trajectory $b$) is long enough to allow the blue mutation to arrive first.
}
\label{fig:delicatesituation}
\end{center}
\end{figure}

However, this model is not quite completely specified:
unlike in the constant-speed case, the collision of two waves traveling at different speeds presents a delicate situation.
Even laying aside questions of how the waves interact (does their interface grow more quickly?), 
the resulting patterns are not described by the simple analogue of equation \eqref{eqn:Adefn},
essentially because if waves are accelerating, then one mutation may surround another.  The problem is depicted in Figure \ref{fig:delicatesituation}.
In defining the regions $A_i$ associated with each type $i$ in the original way, 
we say that each mutation will occupy by time $t$ any previously unoccupied points no more than distance $f(t)$ away.
However, as is shown in the last time slice of Figure \ref{fig:delicatesituation}, 
this requires the path leading to the occupation of certain points to {\em pass through} regions already occupied by another type.
In the constant-speed case, it was always clear which points were reached first by each mutation (for a proof, see \citet{moller92}).
A natural definition of the regions $A_i$ that jibes with the ``wave-like'' intuition would be that at time $t$, 
region $i$ expands outwards (perpendicularly to its boundary) at rate $f'(t-t_i)$, if it is not blocked by another region.
This definition would allocate the grey region in Figure \ref{fig:delicatesituation} to the (blue) mutation that arose second.
Unfortunately, this model seems significantly more difficult to analyze, due to possible interactions between many mutational origins.
Both definitions are likely equally good approximations to the true dynamics,
which are inherently stochastic, especially in the accelerating wave case \citep{mollison1972spatial}.
Furthermore, if allowing mutations to interfere with each other only slows the waves down,
the original definition will be more conservative than another that allows interference, in that it results in strictly fewer independent mutational origins.

With this in mind, we stick with the simple analogue of equation \eqref{eqn:Adefn}, defining formally for $t>t_i$
\begin{equation} \label{eqn:generalAdefn}
    A_i(t) = \{ x \in U : \; f^{-1}(\|x-x_i\|) - (t-t_i) \le \min\left(0,f^{-1}(\|x-x_j\|) - (t-t_j)\right) \; \forall j \} .
\end{equation}
This preserves the same intuition as before, with the modification that waves may now move through each other invisibly
to claim an area on the other side of a distinct region,
which is unrealistic, but results in an underestimate of the number of independent parallel mutations.

Although this more general model has also been studied in the context of crystallization,
where it is generally known as ``Kolmogorov-Johnson-Mehl-Avrami'' dynamics \citep{fanfoni-tomellini},
the focus is on phase transitions and how the proportion of occupied space increases with time (governed by the Avrami equation).
The statistics of the number and shape of the regions in this more general setting seem to have so far been unaddressed.

\begin{figure}[ht]
\begin{center}
\includegraphics[width=1.0\textwidth]{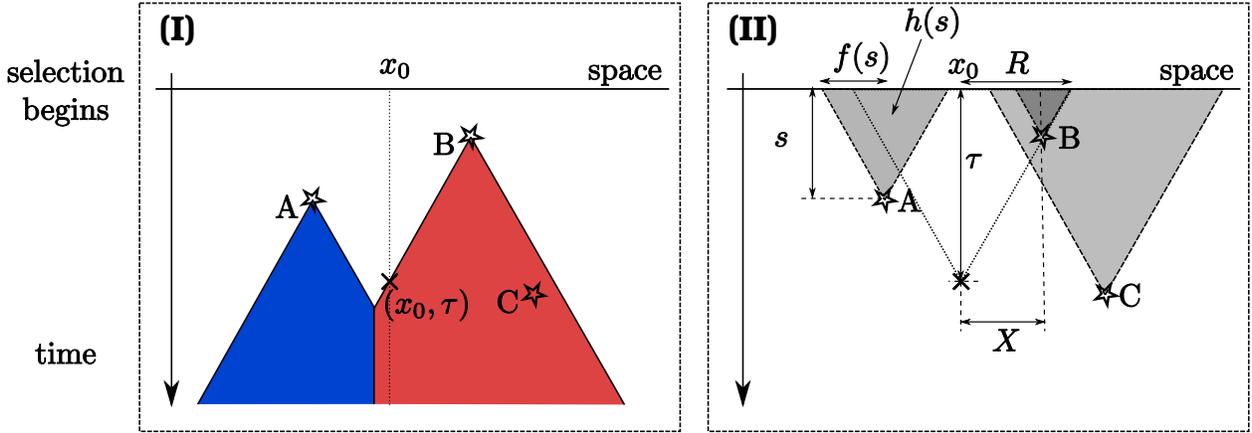}
\caption{ %
{\bf Cones in space-time.}  A single 1D example shown as a space--time diagram.
{\bf (I)} Mutations arise and spread forward in time.
{\bf (II)} Mutations can only spread if a cone backward in time is free from other mutations.
Mutational origins are marked with stars; $A$ and $B$ are successful.
Mutation $A$ arises at time $s$; 
the probability that it is the first at that point depends on the area of the cone stretching backwards from it (shaded);
this cone has base $f(s)$ and area $h(s)$.
Mutation $C$ does not spread because it occurs where $B$ has already reached; 
we can see this because point $B$ lies in the cone stretching backwards from $C$.
The point $x_0$ is first occupied at time $\tau$;
two quantities of interest are $X$, the distance from $x_0$ to the origin of the mutation that eventually encloses it,
and $R = f(\tau)$.
Note that the ``cones'' will not generally have straight sides unless $f(t)=vt$.
}
\label{fig:exclusions}
\end{center}
\end{figure}

Any quantity we might be interested in follows in principle from a calculation
with Poisson point processes--- see \citet{kingman-poisson-processes} or \citet{daley-vere-jonesI} for background.
Figure \ref{fig:exclusions} depicts the general procedure; it will be useful in viusalizing the following argument.
For instance, fix a time $t$ and a point $x_0$ that is at least distance $f(t_0)$ from the boundary.
The location $x_0$ is covered by a mutant type if at some time previously, 
a mutation arose nearby enough that it could have reached that location in the time elapsed since it arose.
If the mutation arose $t$ time units in the past, it must be within distance $f(t)$.  
Therefore, to see if a point $x_0$ has {\em not} yet been covered at time $t_0$, we need only look back through time to see if,
at each time $t$ units in the past, the corresponding circle with radius $f(t)$ is empty of new mutations.
A trajectory of radially expanding circles moving back through time sweeps out a cone in space-time 
(whose sides are not straight if the speed is not constant);
we denote by $h(t)$ the area (in space-time) of such a cone of height $t$,
defined by $h(t) = \left| \{ (s,y) : s \ge 0 \; \mbox{and} \; \|x-y\| \le f(t-s) \} \right|$.  
The area of this cone multiplied by the population density represents the total number of individuals a mutation could have arisen in.
Since successful mutations form a Poisson point process in space and time,
we know that the number of successful mutations in such a cone is Poisson distributed with mean $\lambda h(t)$.

This simple fact gives us the distribution of the time until some (any) adapted type arrives at a given location.
Denote by $\tau$ the time that $x_0$ is first reached by an adaptation, shown in Figure \ref{fig:exclusions}.
(Formally, $\tau=\inf\{t>0:x_0~\in~\bigcup_i~A_i(t)\}$.)
Then $\tau>t$ if the cone with base at $(x,t)$ is empty of successful mutations, and so
\[
    \P \{ \tau > t \} = \exp \left( - \lambda h(t) \right) ,
\]
and so if the species range $U$ is infinite (we mean $\R^d$), the expected time until $x_0$ is adapted is
\[
    \E[\tau] = \int_0^\infty \exp( - \lambda h(t) ) dt .
\]

The entire process is parameterized by three things: the wave expansion profile $f(t)$, the mutation intensity $\lambda$, and the region $U$.
However, by changing variables as before, we can remove $\lambda$ and the scaling of $U$:
the process characterized by $(f(t),\lambda,U)$ is equivalent
up to a linear scaling of time and space, to the process characterized by $(f(t/\lambda x^d)/x, 1, U/x)$.
($U/x$ is the region $U$ rescaled by the numeric factor $x$.)
In the constant speed case $f(t)=vt$, then we are left with a single parameter: 
$f(t/\lambda x^d)/x = x^{-(d+1)} vt/\lambda = (\chi/x)^{d+1} \omega(d) \; t$,
where $\chi=(v/(\lambda \omega(d)))^{1/(d+1)}$ is the characteristic length of equation \eqref{eqn:constchidefn},
and recall that $\omega(d)$ is the area of the sphere with radius $1$ in $d$ dimensions.
This suggests defining the characteristic length $\chi$ more generally to satisfy the equation 
\begin{equation}  \label{eqn:chidefn}
	f\left(\frac1{\lambda \chi^d \omega(d)}\right)=\chi .
\end{equation}
This has a natural interpretation.  
An unobstructed mutation in $t$ units of time will cover an area with radius $f(t)$.
The expected number of other mutations that occur in that area over that time period is $\lambda \omega(d) t f(t)^d$.
After the wave has traveled distance $\chi$, this expected number is $\lambda \omega(d) f^{-1}(\chi) \chi^d = 1$.
Thus, $\chi$ is the distance traveled by an unobstructed spreading wave
before it is expected that one other successful mutation would have arisen within the area enclosed so far.

Regardless of the wave expansion profile,
if the range is large relative to $\chi$, there will be many mutations with high probability;
and conversely, a range small relative to $\chi$ will have few mutations with high probability.
Indeed, by the time a single wave has traveled distance $\chi$, in any other circle of radius $\chi$
there is a good chance that another mutation has already occurred,
so that the chance a single mutation manages to fix everywhere before another arises is small.

\subsubsection{Global properties}
Now we derive formulas for a few other quantities of interest:
the mean density of successful mutations; 
the expected area covered by a typical mutation;
and properties of the distance from a chosen point to its mutational origin.
We provide these results and their derivations to give more intuition for the Poisson point process and space-time cones.
M{\o}ller treated only the constant speed case but allowed inhomogeneity in time;
it seems that the case of a general wave expansion profile $f$ does not appear in the literature.
Less mathematically inclined readers may at this point skip to Section \ref{ss:fattailed}, or even Section \ref{ss:simulations},
without much loss of continuity.

We assume that the region $U$ is $\R^d$.
In general, the shape of the region will have some effect on these quantities,
but if the region is large in all directions relative to $\chi$ then the effect will be small.
Also, without loss of generality, we can now rescale time so that the mutational flux $\lambda=1$ 
(which also affects the wave expansion profile $f$).
When we compute numerical examples, we'll need to remember that time is in units of $1/\lambda$ generations.

We will make much use of the volume of the ``cone'' in space-time swept out by the expanding mutation over $t$ time units,
defined to be $h(t) = \int_0^t \omega(d) f(u)^d du$,
and depicted in Figure \ref{fig:exclusions}.
We will also want to know the volume of the cone with radius $r$ at the base,
which (abusing notation a bit) we define by $h(r) = \int_0^{f^{-1}(r)} \omega(d) f(u)^d du$.

First consider $\nu$, the mean number of successful mutations per unit area.
Let $g(t,x)$ be the probability that a successful mutation arose at location $x$ and time $t$, and that no other mutation reached $x$ earlier.
The mean number of successful mutations originating within a region $W$ is the integral of $g(t,x)$ over $[0,\infty)\times W$.
The probability that some mutation arose in a small region of size $\epsilon$ about $(t,x)$ is approximately $\epsilon$;
and the probability that no other mutation had already reached that point is $\exp(-h(t))$.
Since this does not depend on $x$, the mean number of successful mutations originating in $W$ is equal to the area of $W$
multiplied by $\nu$, which we now know to be
\begin{equation} \label{eqn:generaldensity}
    \nu = \int_0^\infty \exp( - h(t) ) dt .
\end{equation}
A subregion $W$ of total area $|W|$ will have on average $\nu |W|$ successful mutations arising within it 
(although note that mutations not arising within $W$ could invade).

Now consider the area finally occupied by a ``typical'' successful mutation, which we denote by $A$. (Technically, with distribution given by the Palm measure.)
Since the total area of a region divided by the number of mutations in the region, which converges to $\E[A]$ as the size of the region increases, 
and this can also be seen to converge to $1/\nu$,
we also now know that $\E[A] = 1/\nu = \left( \int_0^\infty \exp( - h(t) ) dt \right)^{-1}$.

Finally, fix some point $x_0$.
Let $X$ be the distance from $x_0$ to the origin of the mutation that eventually encloses it, let $\tau$ be the time until $x_0$ is first covered, and let $R = f(\tau)$
(depicted in Figure \ref{fig:exclusions}).
As in the constant-speed case, the probability that $x_0$ has not yet been reached by time $t$ is $\exp(-h(t))$,
and so the distribution of $\tau$ is given by $\P\{ \tau > t \} = \exp(-h(t))$.
Then to have $X=x$ and $R=r$, we need a mutation to have arisen at the appropriate time somewhere in the ring of radius $x$ about $x_0$,
and no other mutation to have already arisen in the cone whose point is at $(\tau, x_0)$.
The cone is empty with probability $\exp(-h(r))$,
and since the ring of radius $x_0$ and width $dx$ has area $d \; \omega(d) x^{d-1} dx$ (and the number of points in it is Poisson distributed),
the contribution to the joint probability density of $X$ and $R$ is $d \; \omega(d) x^{d-1}$ (note that each $d$ appearing in this last expression denotes the dimension).
This joint density is then
\begin{equation}
   \P \{ X \in dx, R \in dr \}
       = \exp\left( - h(r) \right) d \omega(d) x^{d-1} dx dr ,
         \qquad 0\le x \le r .
\end{equation}
Integrating over possible values of $R$, we get a density for $X$:
\begin{equation}
\P\{ X \in dx \} = d \omega(d) x^{d-1} dx \int_x^\infty e^{-h(r)} dr
\end{equation}
Changing the order of integration, the moments of $X$ can be written
\begin{equation}
  \begin{split}
    \E[X^n] &= d \omega(d) \int_0^\infty e^{-h(r)} \int_0^r x^{n+d-1} dx dr \\
    &= \frac{\omega(d) d}{n+d} \int_0^\infty r^{n+d} e^{-h(r)} dr .
  \end{split}
\end{equation}

\subsection{Stochastic waves} 
\label{ss:stochasticwaves}
We have treated the wave expansion as deterministic,
but in real biological systems, the true dynamics will be stochastic.
Established theory for Poisson processes allow us to at least write down analogous expressions
in the general, stochastic case.  
For example, if the selection coefficient varies over a sufficiently small range that mutational exclusion approximately holds,
we could model each type as having a randomly chosen speed.
Alternatively, the shape itself could be random, 
with long-distance migrants causing discrete patches to appear outside of the main spread of the wave in a stochastic manner.
Happily, it turns out that if we make the same definition as at the start of Section \ref{ss:generalspeed}
to avoid the ``delicate situation,''
then the equations for $\nu$, the distribution of $\tau$, and $\E[A]$ all still hold,
after replacing $h(t)$ by the mean volume swept out over $t$ time units,
as we prove in Appendix \ref{apx:stochastic}.

\subsection{Some fat-tailed kernels} \label{ss:fattailed}

Dispersal kernels that lead to accelerating waves are thought to be important for the spread of real organisms \citep{shigesada1997invasions}.
A consequence of the existence of a characteristic length is that the long-time behaviour of a wave has little effect---
most waves spread no farther than a small multiple of $\chi$,
and hence different $f$ that agree over the scale of $\chi$ will result in similar patterns.
Since even exponentially bounded kernels 
can lead to waves that accelerate for some time before reaching the asymptotically constant speed,
it is of interest to look at different wave expansion profiles,
regardless of what we believe about the tail of the dispersal kernel.


\cite{kot96} divide the class of fat-tailed distributions into those distributions with finite moments of all orders,
and those without.
Since our interest is in the resulting patterns of diversity,
and it is not our intention to analyze the precise behavior of the wave under different models of dispersal,
in Appendix \ref{apx:fourierspeeds} we will follow Fourier transform methods of \cite{kot96} 
to somewhat heuristically obtain expressions for the wave expansion profile $f(\cdot)$
for two families of fat-tailed distributions without moment generating functions:
the stretched exponentials (with moments),
and the L\'evy symmetric stable distributions (without moments).
We then apply the theory of Section \ref{ss:generalspeed}.
These results are used in Section \ref{ss:numerics} to compare the three families
at some biologically relevant values.

\begin{figure}[ht]
\begin{center}
\includegraphics[width=1.0\textwidth]{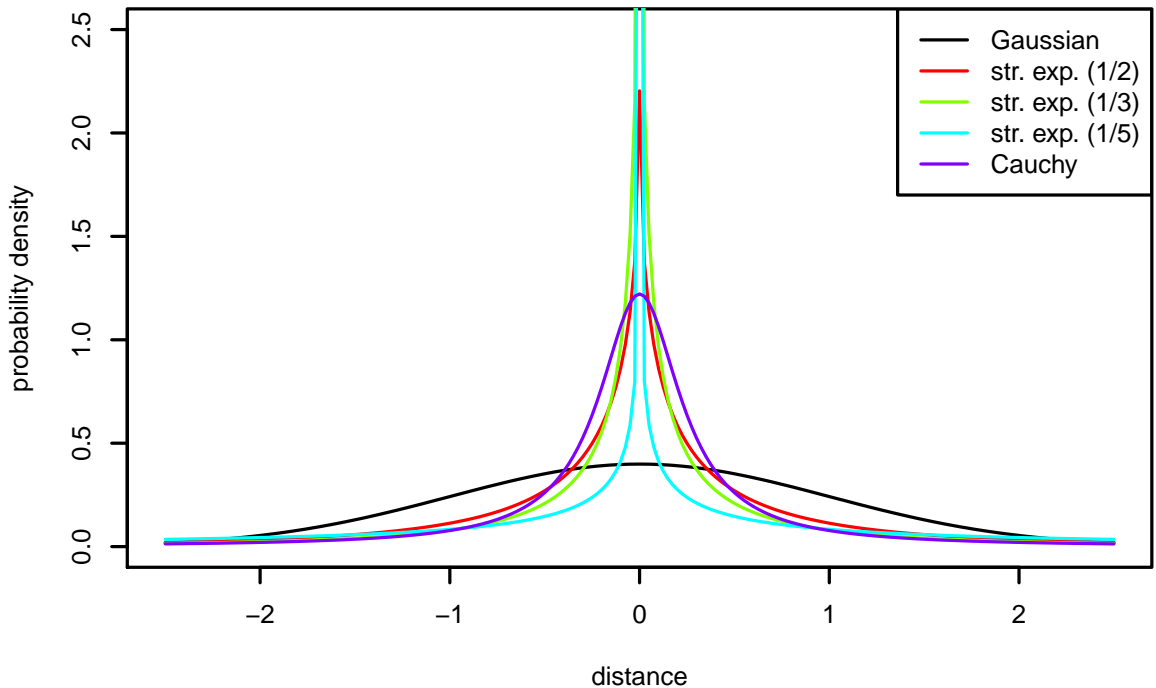}
\caption{ %
{\bf Probability densities for several kernels.} Shown are a standard Gaussian, stretched exponentials with $\alpha$ values $\frac12$, $\frac13$, and $\frac15$, and the Cauchy distribution (symmetric stable with $\alpha=1$). Each is scaled to match the $75^\mathrm{th}$ quantile of the standard Gaussian.}
\label{fig:distributions}
\end{center}
\end{figure}

All kernels have a scaling parameter, denoted by $\sigma$, which is the scale on which distance is measured.
To then compare different kernels, we need to standardize them to each other;
however, we can't match standard deviations because for some of these kernels, the standard deviation is not defined.
We have chosen to standardize so that the interquartile ranges 
({\em IQR}, the difference between the $75^\mathrm{th}$ and the $25^\mathrm{th}$ quantiles, equivalent here to the {\em median absolute deviation})
match those of the standard Gaussian.  See Figure \ref{fig:distributions} for a depiction of the resulting kernels.

\section{Results}

\subsection{Simulations} \label{ss:simulations}

\begin{figure}[ht]
\begin{center}
\includegraphics[width=1.0\textwidth]{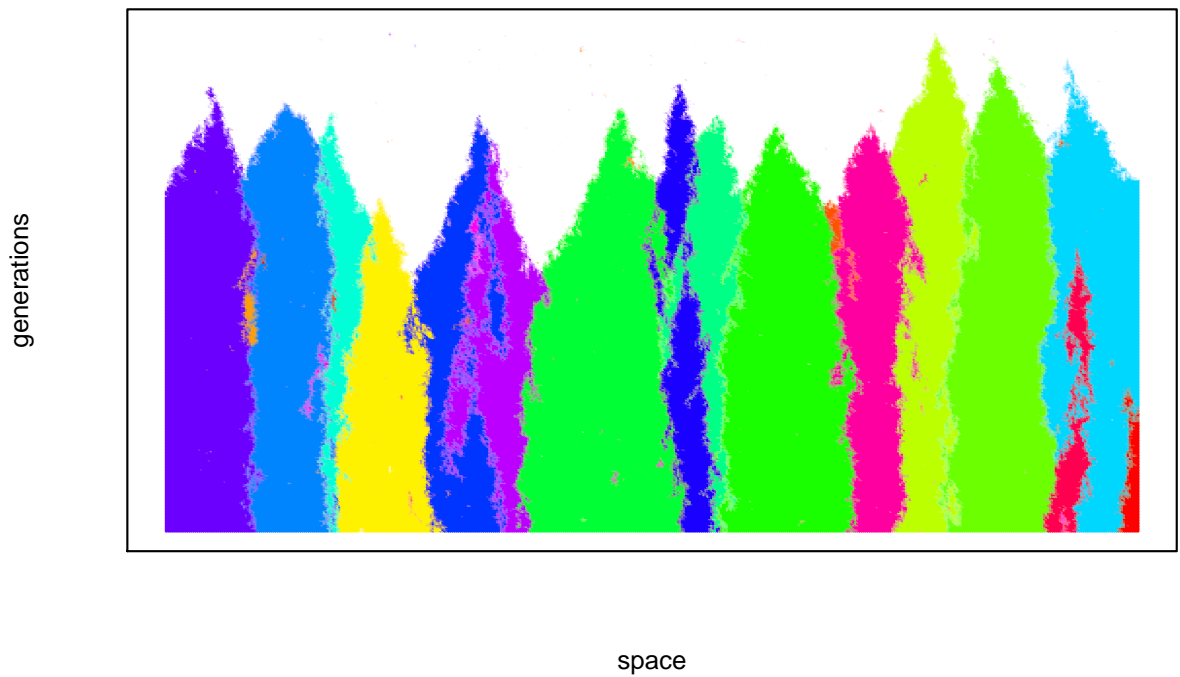}
\caption{  %
A space-time plot of a single run of a simulation on a linear array of 500 demes each of size $100$ over 20,000 generations.  
The parameters were $s=0.1$, $m=.01$, $\mu=4\times10^{-6}$, and migration was nearest-neighbor.  
Time runs down the page; different colors label different types, and areas occupied by more than one type are colored by a mixture of the colors
(local drift is strong in this simulation, so most demes have only one type).
Each distinct ``cone'' has a unique type despite similarities in color choice.  
Note that types expand at roughly constant speed until encountering another type, and that mixing, while present, happens on a longer timescale.
Types that appear where the advantageous type is already fixed (e.g.\ the orange bit between the purple and blue regions on the left) are unlikely to survive, even if they locally escape drift.
\label{fig:1dsim2}
}
\end{center}
\end{figure}

\begin{figure}[ht]
\begin{center}
\includegraphics[width=1.0\textwidth]{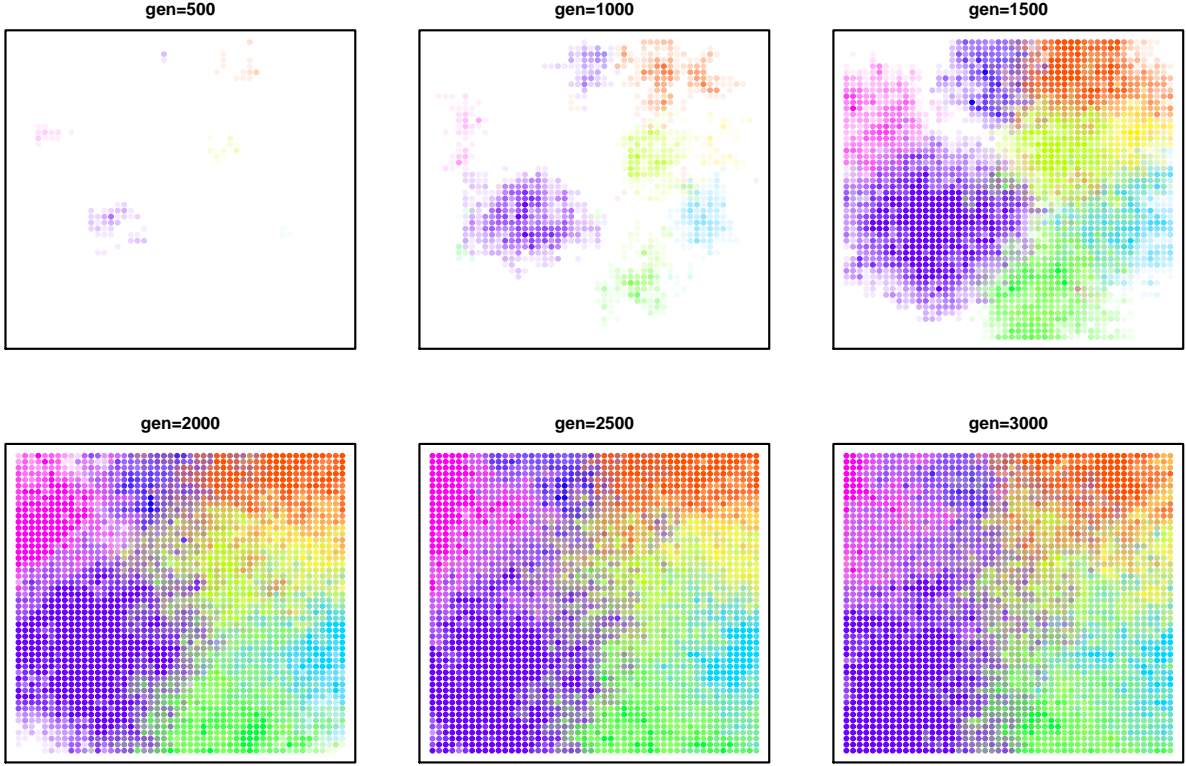}
\caption{  %
\label{fig:2dsim}
{\bf Six time slices of an example simulation in a two dimensional range,}
showing initial establishment and expansion of types, and the beginning of mixing
(which happens much slower than expansion).
The population was composed of a $60\times60$ grid of demes with $1000$ individuals in each.
Different colors correspond to different types, and white is the ancestral type;
when more than one color occupies a deme, the colors are mixed,
so that eventually, if all colors spread to all demes, the entire population will be grey.
}
\end{center}
\end{figure}

To test the robustness of the theory to deviations from the assumptions,
we implemented simulations in {\tt R} ({www.r-project.org}).
In the simulations, the population is a rectangular grid of demes with $N$ haploid individuals in each.
Each generation, each individual independently produces either one or no offspring;
the probability she produces an offspring is $r$ if she is of the ancestral type,
and it is $r(1+s)$ if she is of any mutant type.
Each offspring is a new, as--yet--unseen type with probability $\mu$; otherwise it is the same type as the parent.
Next, all individuals have the chance to migrate, which they do independently of each other with probability $m$ to a nearest-neighbor deme.
We also used a truncated power-law dispersal kernel, which gave similar results, which we do not display.
Those that migrate outside the population are lost.
Finally, the demes are resampled down to size $N$.
As in the Wright-Fisher model, death does not occur explicitly, but only during the resampling phase.
The history of a typical run on a one-dimensional grid is shown in Figure \ref{fig:1dsim2} with time in the vertical direction,
and several time slices of a typical two-dimensional simulation are shown in Figure \ref{fig:2dsim}.

The simulations use discrete generations and discretized space,
in contrast to our theory, and so provides better support for the approximations we use.
The discrete model is, however, in the domain of attraction of the classical wave of advance---
if we rescale time so each generation lasts $\epsilon$, rescale space so the grid spacing is $\sqrt{\epsilon}$,
and make selection weak, setting $s=s'\epsilon$,
then in the limit as $\epsilon \to 0$ and $N \to \infty$ in the appropriate manner,
we expect the frequency of mutant types to satisfy the Fisher-KPP equation \eqref{eqn:FKPP}
with $s=s'$ and $\sigma^2 = m 2^{-(d-1)}$.


\begin{figure}[ht]
\begin{center}
\includegraphics[width=1.0\textwidth]{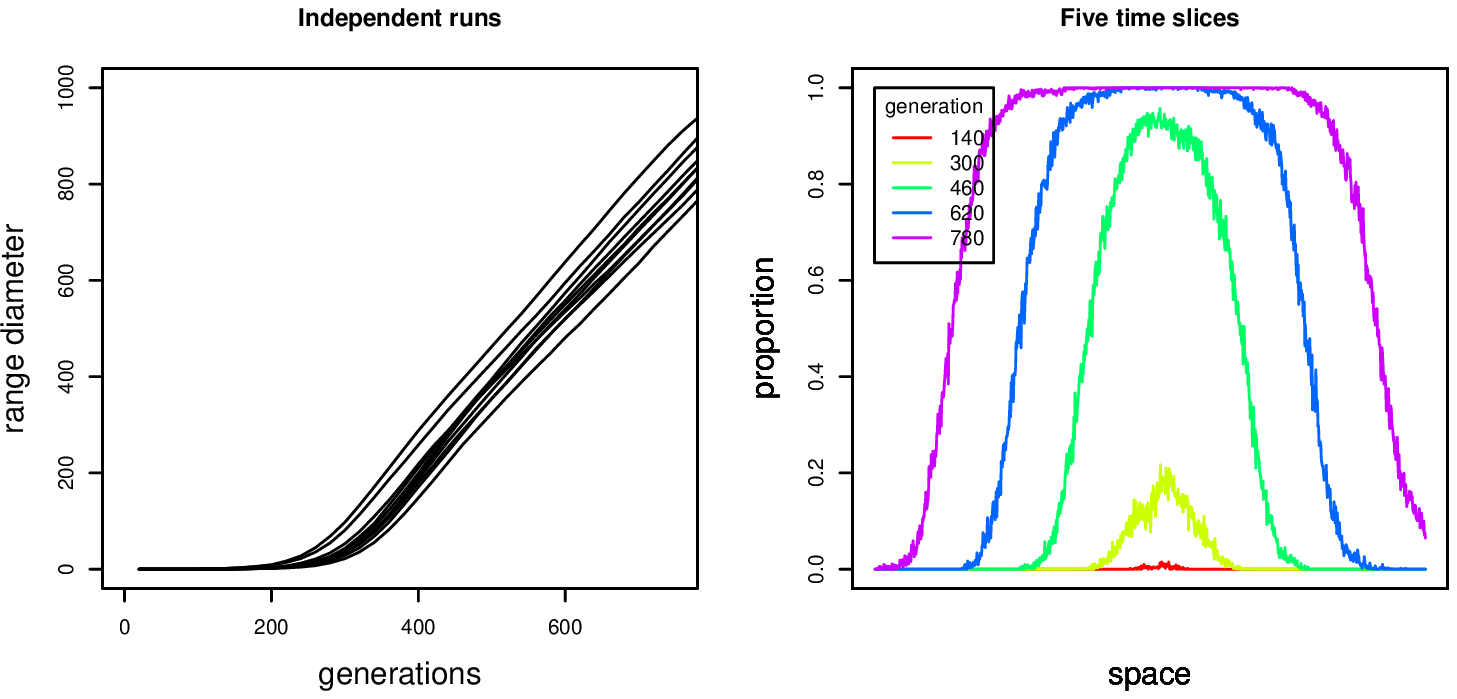}
\caption{  %
\label{fig:takeoff}
{\bf Simulations show constant-speed waves.} 
Typical spread of a wave in a 1D simulation, showing quick settling down to a constant speed.
On the left are plots of the range diameter through time in ten independent realizations;
on the right are five time slices of a single realization.
In each case, the mutation rate was set to zero and a mutant was seeded in the center of the population
to start the wave.
}
\end{center}
\end{figure}

We first used simulations to investigate how well the wave behavior predicted by the Fisher-KPP equation approximates the wave speed of this discrete model.
Deviations in the wave speed and the wavelike form of the spread are likely to be the largest sources of error in our approximations.
The results from a representative set of simulations are shown in Figure \ref{fig:takeoff},
performed on a linear grid of 1000 populations, with $s=.1$, $N=1000$, $r=.4$, nearest-neighbor dispersal, $\mu=0$,
and an initial seed of 20 mutant individuals at the origin.
For local dispersal the wave speed was constant, after a short transient period of random length,
as the trajectories on the left of Figure \ref{fig:takeoff} show.
The speed found was 1--2 times faster than predicted,
likely because our simulations have discrete generations ($r$ was typically between $\frac{1}{10}$ and $\frac12$)
and discrete demes,
rather than the continuous space and time of the Fisher-KPP equation.
However, the wave speed does depend linearly on $m$ and $\sqrt{s}$ as predicted across different dispersal distributions (results not shown), 
which we view as the important verification.

\begin{figure}[ht]
\begin{center}
\includegraphics[width=1.0\textwidth]{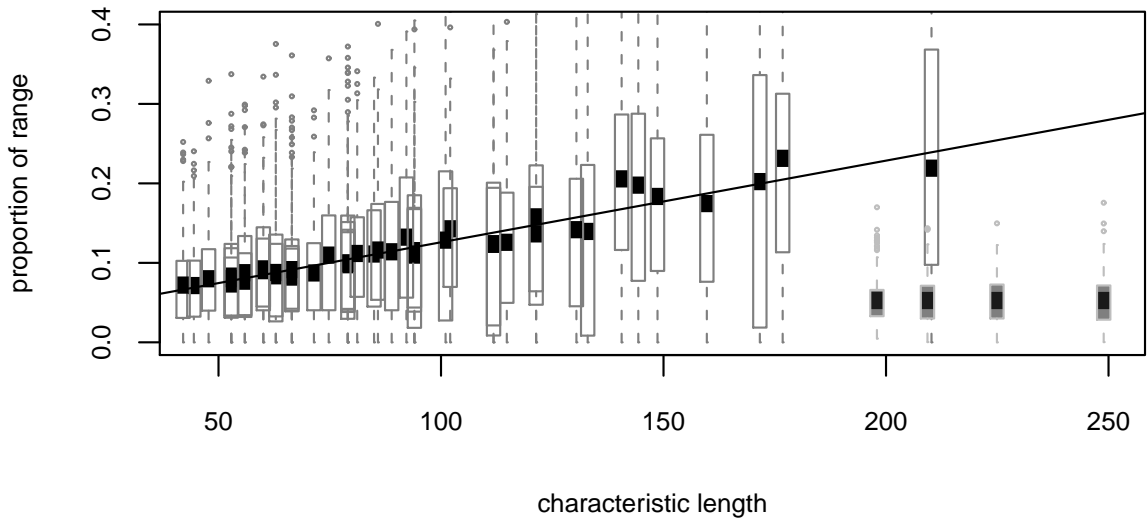}
\caption{  %
\label{fig:empiricalsizes}
{\bf Empirical mean range sizes occupied by distinct types depends linearly on the characteristic length.}
The results of 818 simulations at 45 distinct parameter combinations of $N$, $m$, $r$, and $s$ and local dispersal (unshaded boxes)
as well as 90 simulations at a larger value of $N$ and with dispersal on the order of the species range (shaded, grey boxes).
All had $\mu$ fixed at $10^{-7}$; see text for parameter choices otherwise.
For each set of parameter values, between 10 and 30 independent simulations were carried out,
and the areas (as a proportion of the total range) occupied by each distinct type were tabulated.
A boxplot of the resulting 45 distributions is shown, 
with the characteristic length computed using Equation \eqref{eqn:constchidefn} on the horizontal axis.
Since the characteristic length does not depend on $r$, some boxplots overlap.  
The boxplots are standard (boxes extend from the first to the third quartiles), 
except that the {\em means} are shown as a black box,
since it is the mean occupied area which is expected to be linearly proportional to $\chi$.
Shown is the regression line of proportion occupied against characteristic length, using only simulations with local dispersal.
}
\end{center}
\end{figure}

To test our prediction that the mean radius of the regions occupied by distinct mutations was captured by our compound parameter, the characteristic length,
we also compared the size of regions occupied by each type in simulations to the theoretical predictions.
We ran 818 simulations on linear grids of 1000 demes with nearest-neighbor migration at 45 different combinations of the following parameter values
chosen to obtain a good spread of characteristic lengths:
migration probability $m \in (0.025,0.05,0.1,0.2)$, reproduction rate $r \in (0.1,0.2,.05)$, local population size $N\in(200,600,1000)$,
and selection coefficient $s \in (0.08,0.16,0.24,0.32,0.4)$.
Simulations were stopped at the first time the ancestral type went extinct, 
and the proportions of the total population that were of each type was computed.
In a one-dimensional range,
we expect the mean area occupied by a type (and hence, proportion of the total population, since all simulations had the same number of demes)
to increase linearly with the characteristic length,
with a slope depending on the details of the model (which we did not attempt to compute).
The resulting distributions of proportions are shown as a function of characteristic length in Figure \ref{fig:empiricalsizes} (unshaded boxplots),
showing a good linear fit of the mean area to the characteristic length (solid line),
confirming that our theoretical predictions fit the simulated model well.

\paragraph{Limits of the model} \label{ss:limits}
In our view one of the more serious approximations we make is that once the allele is introduced to a deme it quickly reaches its equilibrium frequency locally. 
This allows us to assume that the time delay between when the selected allele arises and when it begins to spread 
as a wave of known speed is short and relatively constant across alleles.
This approximation is important because it underlies our assumption of allelic exclusion, 
i.e.\ that in areas reached by a spreading mutation, subsequent mutations do not arise and escape drift fast enough to also spread.
In both cases ``fast'' must be quicker than the time scale on which mutations arise and escape drift locally.
To demonstrate the shortcomings of this approximation, 
we ran additional 90 simulations on the same linear grid
with a Gaussian dispersal kernel whose SD is $1/10$ the range size (100 demes), chosen to violate this assumption.
The results are shown in Figure \ref{fig:empiricalsizes} (shaded boxplots).
The local population size $N$ was set to $10,000$, to make the characteristic lengths comparable, 
and the other parameters were a subset of those chosen above ($m=0.05$, $r=0.1$, and $s \in (0.16, 0.24, 0.32, 0.40)$).
The observed proportion of the range occupied per type is far lower than would be predicted from our characteristic length,
which is expected if mutational exclusion does not hold.
Indeed, for these examples the width of the wave (which is approximately $\sigma/\sqrt{s}$ \citep{fisher1937wave,KPP1937})
is around $1/3$ the characteristic length, 
indicating that a significant number of successful mutations will arise in a location where another type has already begun to occupy.
In fact, we see in Figure \ref{fig:empiricalsizes} that 
for these simulations the mean number of types does not appear to depend on $s$, as expected in the panmictic model of \citet{softsweepsII}.
If the width of the wave was much smaller than the characteristic length, this would not be a problem.
We return to the question of when our approximations hold in the Discussion.


\subsection{Biological parameters and the characteristic length} \label{ss:continuousresults}

The best summary of the probability and spatial scale of parallel adaptation in our model is the characteristic length.
The simple form of the characteristic length, especially in the classical Fisher-KPP case, 
allows us to find how the various parameters affect the spatial scale and probability of parallel adaptation.  
For example, doubling the local effective population density has a similar effect to halving the standard deviation of the dispersal distance. 
Intuitively, this reflects the fact that halving the standard deviation of the dispersal kernel doubles the time it takes a mutation to spread a given distance, 
which in terms of the chance of parallel adaptation is equivalent to doubling the population density. 
Note, however, that these two changes are not equivalent in terms of the time it takes adaptive alleles to spread throughout the entire range---
doubling the density will decrease the time until adaptation, while halving dispersal distances will increase the time,
as seen in Equation \eqref{eqn:etau}.
Furthermore, it is intuitively clear--- and this can be made rigorous--- that in a region large relative to the characteristic length,
many parallel mutations are very likely. 

With this in mind, we here compute characteristic lengths at some representative parameter ranges and in some specific situations,
to give a sense of under what conditions parallel adaptation is likely.
Consider a population spread over an area the size of Europe, here defined as the area west of the Ural mountains, 
which has an area of about $10^7 \mathrm{km}^2$ (about a third the area of Africa) and is about 4,000 km across in the longest direction. 
Therefore, if $\chi < 4000\;\mathrm{km}$, multiple mutations are likely.
We take two population densities, $\rho = 2$ and $\rho = 0.002$, chosen respectively to reflect: 
the human population density 1000 BCE \citep{Atlas:popdensity}; 
and the long-term ancestral effective population size of humans, $N_e=10000$, spread out over the area of Europe. We vary the mutation rate between $10^{-8}$ and $10^{-4}$ mutations per generation, representing a mutational target of between $1$ and $10,000$ base pairs (the typical mutation rate in humans is $\sim 10^{-8}$ per base per generation).
We keep the selection coefficient $s=0.01$ fixed, since the results are fairly insensitive to changes in $s$. We vary the typical dispersal distance $\sigma$ between one and one hundred kilometers (in line with human dispersal estimates \citep{Wijsman:1984}). Finally, we compare several different dispersal kernels--- rescaled so that each shares a common interquartile range---
to see how fatness of tails affects the results (the stretched exponential, with $\alpha=1/2$, matches that used by \citet{Novembre2005}).

\begin{figure}[ht]
\begin{center}
\includegraphics[width=1.0\textwidth]{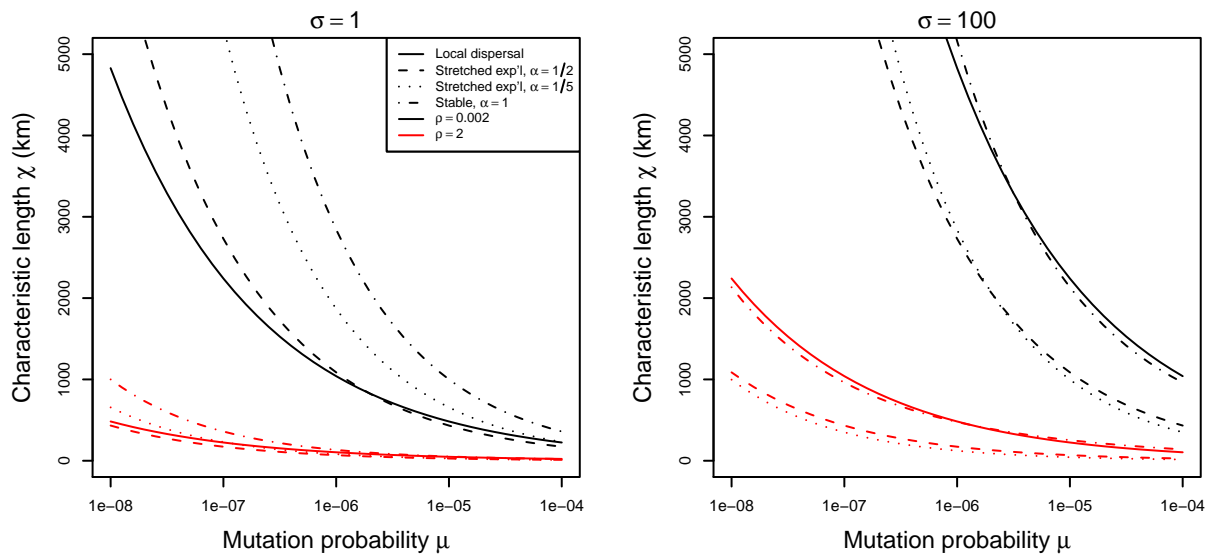}
\caption{  %
{\bf Characteristic lengths under various models at different parameter values.}
If the species range is large relative to the characteristic length, parallel adaptation will occur with high probability.
In each plot, the per-individual mutation rate is along the horizontal axis.
The two plots have the scale parameter $\sigma$ of each dispersal kernel set at 1km and 100km, respectively.
Line type refers to different dispersal kernels: 
Gaussian (or short-range) dispersal, 
stretched exponential with $\alpha=\frac12$ and $\frac15$, 
and Cauchy (stable with $\alpha=1$).
In each case the distributions were normalized to match at the 95th quantile.
Color denotes different values of $\rho$, 
taken to be two possible values for the appropriate density of the ancestral human population in Europe (see text).
\label{fig:xibymut} 
}
\end{center}
\end{figure}

From Figure \ref{fig:xibymut}, we see that the expected degree of parallel adaptation depends strongly on the parameters.
At the lower population density (black lines), it is not until mutation rate is on the order of hundreds or thousands of base pairs (depending on $\sigma$)
that independent origins are likely;
at smaller mutation rates parallel adaptation is highly unlikely.
At the higher population density (red lines),
independent origins are likely even if the mutational target is only a few base pairs; 
at larger mutation rates the characteristic length falls to only tens or hundreds of kilometers,
indicating a ubiquity of mutational origins.
This is to be expected, because with $\rho=2$, an area the size Europe would have $10^7$ individuals, thus a mutation rate of $10^{-4}$ per generation,  would produce 1,000 mutant alleles in a single generation. Of course, a characteristic length that is on the order of the dispersal distance means that the model described here--- 
a smooth circular outward spread of alleles---
is no longer a good approximation to the true dynamics.

It is also clear from Figure \ref{fig:xibymut} that over this range of parameters the choice of dispersal kernel 
does not affect the number of mutations as strongly as do other parameters.
The characteristic length does vary between kernels, but not generally enough to change the conclusion that parallel adaptation is likely or unlikely.
At certain parameter combinations the difference is large, but not over much of the biologically realistic range we have examined.

\begin{figure}[ht]
\begin{center}
\includegraphics[width=1.0\textwidth]{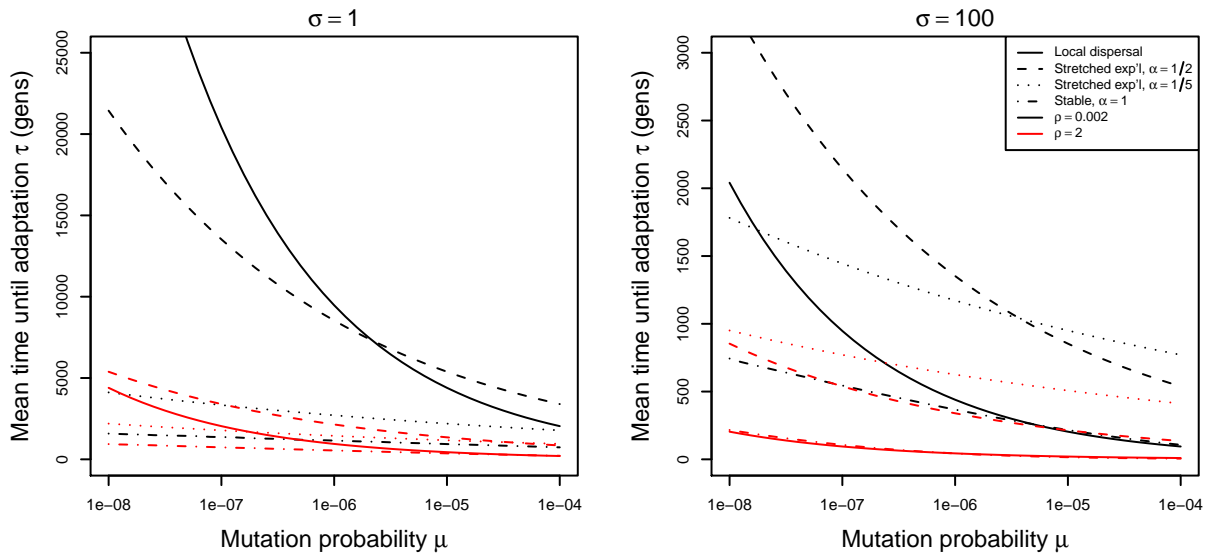}
\caption{  %
{\bf Expected time to the arrival of an adaptive mutation,} in generations, under various models at different parameter values.
Note the different scales on the vertical axes.
In each plot, the per-individual mutation rate is along the horizontal axis,
and other details are as for Figure \ref{fig:xibymut}.
\label{fig:taubymut} 
}
\end{center}
\end{figure}

It is also useful to look at the expected amount of time that the population will take to adapt, 
which we display in Figure \ref{fig:taubymut} over the same set of parameters.
The first observation is that when parallel evolution is likely, it will also take a reasonably small amount of time---
under the conditions where adaptation will take an exceptionally long time, most of the time is spent waiting for a single successful mutation.
The different dispersal kernels, however, lead to more disparate times to adaptation.
Of particular note is the Cauchy kernel, which if dispersal distance is small far outpaces the spread by other dispersal distributions.
As with small characteristic lengths, very small times to adaptation should not be taken too seriously, indicating only that the spread happens quickly.

\subsection{Applications} \label{ss:numerics}

\subsubsection{Why are there so few recent Eurasian-wide sweeps in Humans}
The majority of recently arisen selected alleles, as identified from e.g.\ haplotype patterns, seem to be geographically restricted \citep{Voight:06,Wang:05,Pickrell:09,Coop:09}, occupying broad geographic areas such as Europe or East Asia or Western Eurasia. On the basis of this observation, \citet{Coop:09} and \citet{Pickrell:09} argued that few selected alleles have recently swept to fixation across Eurasia. There are at least three possible explanations for this pattern: 1) there has not been sufficient time for these alleles to spread; 2) the selection pressures are at a local scale, not Eurasia-wide; 3) the selection pressures are shared across Eurasia and different populations have adapted in parallel. These explanations are not mutually exclusive, and to distinguish between them we will need a much more in-depth knowledge of the phenotypes underlying the putative selective sweeps. However, the theory developed here can shed light on whether the last hypothesis is plausible. 

We use a subset of the parameters in the previous section: a Gaussian kernel with $\sigma=100$km, a stretched exponential kernel with $\alpha=1/2$,
and stable distributions with several values of $\alpha$.  
As before, the non-Gaussian kernels are parameterized to have the same interquartile range as the Gaussian kernel with the same $\sigma$.
Table \ref{tab:eurotab} displays our computed characteristic lengths across the above combinations of parameters, at mutation rates of both $10^{-8}$ per generation (a single unique base pair change) and $10^{-5}$ per generation (a $1000$ base pair target, roughly the number of coding bases in a gene). The different dispersal kernels give roughly similar characteristic lengths, suggesting that these numbers are relatively robust to the choice of kernel.

Multiple mutational origins are likely if the characteristic length is shorter than the physical dimensions of the region. Eurasia measures over $8000$km across, and so Table \ref{tab:eurotab} suggests that multiple origins at a single base pair is very unlikely at the lower population density. On the other hand, if the mutational target is large, then multiple origins are likely at low densities, while at high densities independent origins are ubiquitous. 
The complementary cases of $(\rho=2, \mu=10^{-8})$ and $(\rho=0.002, \mu=10^{-5})$ give identical characteristic lengths of about $3000$km, although the time scale on which the mutations spread differs. 
Thus for these two parameter combinations we can expect a few mutations to dominate within continents, and for multiple mutations to be common in a population spread across an area the size of Eurasia. 
Obviously these calculations are very crude, as population densities vary through space and time, and dispersal across continents is not simply a function of geographic distance and individual dispersal. 
Nevertheless, these calculations suggest that it is plausible that for adaptive traits with reasonable mutational targets (e.g.\ a change anywhere within a gene or pathway) even low population densities can lead to parallel adaptation across an area the size of Eurasia, and higher densities almost certainly will.

We note that as human population densities have increased dramatically over time, so too has the probability of parallel adaptation. 
It is interesting therefore to note that a number of recent human adaptations
(e.g.\ sickle cell alleles) involve repeated changes at very small mutational targets in relatively small geographic areas, 
while older adaptations from single changes (e.g.\ skin pigmentation) are more broadly spread. 

\begin{table}[ht]
\begin{center}
  \begin{tabular}{| *{2}l | *{5}r |}
\hline
$\rho$  &  $\mu$      &  Gaussian  &  Stretched  &  Stable 1.25 &      Stable .75 &   Stable .05 \\
\hline                                                                                
2.000       &  $10^{-8}$  &  2240      &  1086       &  1928    &       2360   &       2392    \\
0.002       &  $10^{-8}$  &  22409     &  17224      &  32965   &       39489  &       47124   \\
2.000       &  $10^{-5}$  &  224       &  68         &  237     &       263    &       172     \\
0.002       &  $10^{-5}$  &  2240      &  1086       &  1928    &       2360   &       2392    \\
\hline
\end{tabular}
\caption{ %
\label{tab:eurotab}
Characteristic lengths, in kilometers, at various parameter combinations and with various dispersal distributions.
{\em Stable $\alpha$} refers to a stable dispersal distribution with parameter $\alpha$.
The remaining parameters were: $s=.01$, $r=1$, $\xi=1$, and $\sigma=100\;\mathrm{km}$.
}
\end{center}
\end{table}

\subsubsection{Parallel adaptation of the sickle cell allele}

The sickle cell allele HbS at the betaglobin gene in humans provides a particularly interesting case of putative parallel adaptation. The HbS-allele ($\beta$6 Glu$\rightarrow$Val) has been driven to intermediate frequencies by selection within the past 10,000 years due to increased resistance to malaria of heterozygotes for the allele \citep{Haldane:49,Allison:54,Kwiatkowski:05,Currat:02}. The HbS allele is present on at least four major distinct haplotypes in Africa, each at intermediate frequency within a different geographic region; the haplotypes are named after the population sample where they were first discovered (Central African Republic, Senegal, Benin and Cameroon). This is consistent with multiple origins of this single-base pair change. 
Note that a distinct, malaria resistance allele, HbC ($\beta$6 Glu$\rightarrow$Lys), has also arisen in Africa at the same codon as the HbS allele \citep{Trabuchet:91, Agarwal:00, Wood:05}, increasing our confidence that the mutational input was high enough to allow multiple types to arise. However, \citet{Flint:98} thought the hypothesis of multiple new mutations arising at a single base pair was extremely unlikely, and proposed that it was more likely that gene conversion had spread a single mutation across multiple haplotypes. 
 
The theory we have developed can be used to assess the plausibility of the multiple mutational origins of the sickle cell allele, by exhibiting parameters combinations that yield characteristic lengths consistent with the separation of the sample locations. (Recall that the wave of advance, and thus also our model, works in the case of heterozygote advantage \citep{aronson1975nonlinear}.)
The different HbS haplotypes co-occur within a few thousand kilometers of each other \citep[see Table 5 of ][]{Flint:98},
(noting that these locations are unlikely to reflect the geographic mutational origins, and mutations will have been spread by large population movements). 
As the HbS changes occur at a single base pair, the mutation rate would have been $\sim 10^{-8}$, and we take an $s=.05$ \citep[as in][]{Currat:02}. 
If human dispersal at that time was well approximated by a Gaussian kernel with $\sigma=100\;$km, then a characteristic length of $\sim 1000\;$km would require an effective density of individuals of $\rho \sim 25\;\mathrm{km}^{-2}$, while if $\sigma=10$km, then we would require only $\rho \sim 2.5\;\mathrm{km}^{-2}$. 
This latter set of parameters does not seem unrealistic considering our knowledge of population density and dispersal parameters,
so our model suggests that the hypothesis of multiple origins is not unreasonable.





\section{Discussion} \label{ss:discussion}

We have presented theoretical results on the prevalence of parallel mutations during the sweep of an adapted allele across the geographic range of a species. Parallel adaptation can occur in a species adapting to a shared selection pressure simply because selected alleles may spread slowly enough to allow other mutations to arise and spread elsewhere in the species range. The distribution of the number of unique types is given implicitly by a certain Poisson process, which we summarized by computing the mean values of several important quantities. 
Many features of the continuous model can be captured by a characteristic length, a compound parameter that combines the dispersal parameter, the mutation rate and the population density,
and our simulations confirm that this is a very useful predictor of model behavior.
This characteristic length can be obtained under a wide variety of dispersal kernels, as long as the speed at which selected alleles spread under those kernels can be computed even if that speed is not constant. The regions occupied by distinct types have dimensions on order of the characteristic length, so if the species range is at least as large as the characteristic length, then parallel adaptation is likely. The expected number of parallel mutations is a simple function that, as intuition would predict, decreases with dispersal rate and increases with mutation rate.  
Somewhat counterintuitively,
the results are relatively insensitive to the strength of selection, as selection both hastens the spread of an allele and conversely increases the chances that a new mutation escapes drift. 

\paragraph{Does parallel adaptation require strong population structure?}
Our results confirm the intuitive idea that species with low levels of dispersal (and hence strong geographic genetic structure) may adapt to global selective challenges by parallel adaptation at separated geographic locations. However, the likelihood of parallel adaptation depends on dispersal strength {\em relative to population density} ($\rho/\sigma$), and so geographic parallel adaptation may be an important factor even in species which appear panmictic at neutral markers. 
The observation of relatively little neutral geographic genetic structure merely implies that genetic drift in sub-populations is slow compared to migration, 
which increases with both dispersal distance and local effective population density. 
(Neutral population structure is determined by Wright's neighborhood size, proportional to $\sigma^2 \rho$; see \citet{Charlesworth:03}.)
Increasing dispersal distance ($\sigma$) reduces the chance of parallel adaptation, 
while increasing population density ($\rho$) increases the chance of parallel adaptation. 
Therefore, an absence of strong geographic structure cannot rule out the possibility of geographic parallel adaptation. In this context it may be helpful to consider the global effective population size in a geographically spread population, a common estimator of which is the average coalescent time of pairs of sequences sampled across the species range. Such estimators increase both with lower dispersal distances and higher local effective population densities \citep{Charlesworth:03}. Thus, species with larger global effective population size are more likely to adapt by parallel mutation whether they are truly panmictic (as shown by \cite{softsweepsII}) or selection is dispersal limited, as discussed here.

\paragraph{Signals in patterns of diversity}
While we have discussed our model results in terms of parallel adaptive changes at the same genetic locus, our results do not rely on an assumption of complete linkage between the selected alleles. The base pairs at which these changes occur can be contiguous, partially linked or completely unlinked; we merely require that the mutations are selectively equivalent. Thus our results apply equally well to selected alleles of similar phenotypic effect that have arisen in parallel at different genetic loci, e.g.\ mutations at different genes in the same pathway.  There are a number of potential cases of parallel adaptation within a species where adaptive changes at different genes have produced similar phenotypes in different populations \citep{beach-mice06,beach-mice09,nachman-pocket-mice,hoekstra-pocket-mice}. For example, there is evidence for differences in the genetic basis of the adaptive response to shared selection pressures in European and East Asian human populations \citep{lao-human-pigmentation,norton-human-pigmentation,edwards2010pigmentation}. 

If parallel adaptation has occurred, there are at least two potential signatures in patterns of genetic variation. The first is that, immediately after the sweep, the independent copies of the selected allele will form a spatial patchwork with patches of size $\sim \chi$. Within each patch, a different selected allele will predominate. Each of these mutations may have arisen on a different haplotype, especially if neutral genetic variation varies across the species range. This linked genetic variation will be swept up to high frequency locally and so also form a patchwork, with different haplotypes common within each patch. These patches may maintain sharp boundaries in allele frequencies between them for some time, and so may resemble local adaptation, despite the fact that the causal selection pressures are homogeneous \citep[see ][for discussion]{Coop:09}. This similarity may be further compounded since the boundaries between selected types will tend to occur at geographic barriers to migration, as selected alleles will temporarily be slowed there \citep{Pialek:97}. 

Over time these spatial patterns will be erased through the mixing action of migration. This spatial mixing by migrants will occur in a manner analogous to heat flow, on a time scale given by the diffusive parameter $\sigma$, the standard deviation of the distance between the birth locations of parent and child. In a geographic region of linear dimensions $R$, the patterns will become erased in a time of order $R^2/\sigma^2$ generations. If the alleles have strongly deleterious epistatic (or dominance) interactions then their geographic mixing will be more complicated, and perhaps slowed. A potential example of this is offered by \citet{Williams:05} and \citet{Penman:09} who discuss the role of epistasis in preventing the geographic mixing of  different malaria resistance alleles in humans. Indeed, \cite{Kondrashov:03} has suggested that the relatively slow spread of selected alleles across a species range might allow Dobzhansky-Muller epistatic incompatibilities to arise in parapatry, potentially leading to speciation. 

The second, possibly longer-lasting, pattern is the partition of alleles and their linked haplotypic variation. If genetic drift is slower than spatial mixing, then the initial partition of alleles is spread out uniformly over space long before any alleles disappear or fix through genetic drift.  After mixing by migration, the resulting patterns would be characterized by a region of reduced variation and longer LD surrounding each selected locus. If the independent selected changes occur at the same gene, this will resemble a soft sweep in a single panmictic population as described by \cite{softsweepsII,softsweepsIII}. Indeed, it is possible that some of the characterized putative soft sweeps \citep[e.g.][]{Schlenke:05} arose in this manner. If the changes occur at different genes, there will be a set of partial sweeps at the different loci. Such patterns could potentially explain the apparent excess of partial sweeps, compared to full sweeps, seen in human populations \citep{Coop:09,Pritchard:10}. Thus parallel mutation would allow a population to maintain a higher level of heterozygosity at the selected loci than would sweeps from a single mutational origin.  A related argument has been made by \cite{Goldstein:92}, who discussed the role of genetically redundant mutations and isolation by distance in the maintenance of genetic variation in quantitative genetics models (see also \cite{Lande:91} and \cite{Kelly:06}). 


\paragraph{Exclusion and spatial structure}
Two of our main assumptions--- mutational exclusion and our resolution of the ``delicate situation'' in Section \ref{ss:generalspeed}---
lead to an underestimate of the number of independent, parallel adaptations,
while others--- such as deterministic spread of the waves--- only affect the variance of the number of parallel adaptations.
It is also useful to compare with existing results on panmictic populations.
As described in Section \ref{ss:limits}, simulations using long-distance dispersal that resulted in a nearly panmictic model
produced a much higher number of parallel adaptations than predicted from our theory.
This is in line with the results of \citet{softsweepsII}, 
who have shown that {\em within} a single large randomly-mating population, a high rate of introduction of selectively equivalent mutations 
can allow multiple mutations to escape low frequency before the first to arise fixes in the population.
Extrapolating from their results we see that if population density is high enough in an area 
where a spreading mutation has begun to establish, then other mutations could arise and concurrently spread,
undermining our assumption of allelic exclusion. 
The higher migration is relative to reproductive rate, the closer a model is to the panmictic model of \citet{softsweepsII}.
In general, since it neglects spatial structure, we expect the panmictic model to underestimate the true degree of parallel adaptation as well,
so if the panmictic model predicts more parallel adaptations than does our model, we expect the truth to be closer to the panmictic predictions.
Future work could relax our assumption that the mutations quickly reach equilibrium, 
allowing model predictions more accurate than either model.
In any case, our results provide an underestimate of the prevalence of parallel adaptation, 
but with very widely dispersing species the results of \citep{softsweepsII} may be more appropriate.

\paragraph{Selective equivalence}
Throughout this paper we have assumed that variation in selection coefficient will be much smaller than the strength of selection
(as follows from our focus on {\em parallel} adaptation). 
However, it is unclear how often the {\em strict} selective equivalence holds in practice. 
Mutations that have a convergent effect on a phenotype of interest may differ in their pleiotropic effects,
and even identical changes at the same base pair may have somewhat different effects due to linked variation.

However, the characteristic length only depends weakly on $s$, so the effect of small differences in selection coefficient should be minimal. Further, our results on stochastic waves (Section \ref{ss:stochasticwaves}) suggest that if there are only small differences, it is reasonable to use the mean selection coefficient, and that the mutations will initially form a patchwork with the average size of a patch given by this characteristic length. The time scale over which the patchwork persists will be affected by the differences in selection coefficient. Suppose for simplicity that each newly arising type chooses one of only two distinct selection coefficients, either $s_1$ (a ``weak'' mutation) or $s_2>s_1$ (a ``strong'' mutation) that interact additively. The original patchwork is erased as the stronger mutations push into, or arise within and overtake, areas already occupied by the weak mutations. They do this at speed $\sigma \sqrt{2(s_2-s_1)}$, so the time-scale over which the original tessellation is erased is of order $\chi/(\sigma \sqrt{2(s_2-s_1)})$. The patterns in diversity resulting from multiple types with different selection coefficients will depend on the linkage of the loci underlying the different types. In some cases (e.g.\ full linkage), the stronger allele may push the other out of the population as it spreads; while in other cases (e.g.\ no linkage) the stronger allele will spread throughout the population but not disrupt the spatial pattern of the weaker alleles.




\paragraph{Outlook}
Our results demonstrate that if dispersal is indeed a limiting factor in the spread of selected alleles, then in large geographically spread populations, parallel adaptation will be common. As yet there are relatively few firm examples of parallel adaptive mutations within species, but we believe that this simply reflects the fact that we are just beginning to identify and understand mutations that contribute to adaptive phenotypes. It is notable that many of the cases of geographic parallel adaptation come from humans and the other species where phenotypes that have reasonably well-characterized genetic bases  have been carefully studied (e.g.\  pigmentation, or drug and insecticide resistance). This suggests that further work on other species and phenotypes will uncover many more examples. 

Genes or pathways that harbor different mutations that have swept in non-overlapping parts of the species range will represent good candidates for geographic parallel mutation. One difficulty in interpreting these candidates will come in understanding whether they are approximately genetically redundant with respect to the phenotypes that selection has acted on, or if they dominate in different portions of the species range because they represent locally adapted alleles. The former explanation may be appropriate as a null hypothesis, as it requires fewer differences in selection pressures across the range, and requires only that populations are geographically separated. 

Aside from identifying candidates for geographic parallel mutation, a productive line of research is to understand whether population densities and dispersal patterns are conducive to their occurrence. The spatial density of individuals within a population is likely to fluctuate dramatically over time, 
so the long term effective population size for the species is likely to be a very poor estimate for the rate at which selected mutations arise, especially in populations that have experienced recent rapid growth. The move towards genome-wide population resequencing data will allow the recent effective population size to be estimated from the rare alleles, and the spatial spread of these rare alleles will be informative about recent dispersal parameters \citep[e.g.][]{NovembreSlatkin:09}. 

As yet we know relatively little about the full impact of long-distance dispersal; a situation that will hopefully be improved by the increasing spatial and genomic resolution of population genetic studies \citep[e.g.][]{Novembre:08, Auton:09}, along with the methods to accurately identify subtle signals of gene flow in such data sets \citep[e.g.][]{Price:09}. In many species rare, extremely long distance migrants occur, which can have strong effects on the speed and patchiness with which the wave advances \citep{pacalalewis:2000,clark1998trees,kot96}. While in numerical examples we did not see a strong effect on the likely amount of parallel mutation, this conclusion does not extend to all parameter values, and it is difficult to compare parameters across different dispersal distributions. If migration is not spatially restricted (e.g.\ the fully connected ``island'' model), then we expect the dynamics to be significantly different. There are a number of examples of very rapid spread of selected alleles \citep[e.g.\ in malaria ][]{Wootton:02,Roper:04, Anderson:05}, and of vertically transmitted parasites \citep{kidwell1983evolution, turelli1991rapid}, perhaps suggesting that the extent to which limited dispersal allows parallel adaptation is still a very open question, and likely to vary between species. 
  


\section*{Acknowledgements}
We thank Dave Begun, Chuck Langley, John Novembre, Sebastian Schreiber, Monty Slatkin, and Michael Turelli for helpful discussions. The manuscript benefited from comments on an earlier draft by Chuck Langley, John Novembre and Michael Turelli, as well as the editor and 2 anonymous reviewers. This work was supported by a Sloan Fellowship to GC, and PR was supported by funds from UC Davis to GC and S Schreiber.

\appendix

\section{Stochastic waves}
\label{apx:stochastic}

Here we continue on where Section \ref{ss:stochasticwaves} left off,
to demonstrate when, and in what sense, expressions for the ``mean wave'' suffice,
even if the true behavior is stochastic.

We need to know the probability that an isolated mutation arising at $x$ will first cover a point at $y$ after time $t$,
which we define to be $q(t,r)$, with $r=\|x-y\|$, the distance from $x$ to $y$.
We assume that this quantity only depends on the distance between $x$ and $y$,
regardless of the direction.
This does not require that waves are circular,
only that there are no ``preferred directions''.
We also assume that each wave's stochasticity is independent of the Poisson process of locations and each other.

The mean time until a point $x_0$ is reached by the adaptive type can be found in a manner entirely analogous to the deterministic case.  
Before, all waves were the same, so we only needed to keep track of their origins;
now mutations can be thought of as having different ``types'',
chosen independently and randomly from a type space $\mathcal{V}$,
so that each point now consists of a triple $(X_i,T_i,V_i)$, 
recording the location $X_i$ and time $T_i$ of the mutational origin, and what type of wave $V_i$ it will have, respectively.
The types $V_i$ are chosen independently from some common distribution.
For a simple example, if only the speed is random, then $\mathcal{V}$ is the set of positive real numbers, 
$V_i$ simply records the speed of the wave, and $q(t,r) = \P\{ V_i > r/t \}$.

Just as before, the number of points in any region of time, space, and type space is Poisson distributed, with mean proportional to its measure.
We now measure ``area'' in the third coordinate using the distribution of $V$,
so that the number of mutations that occur in a spatial region of area $A$,
over a time interval of length $t$, and with type lying in some set $\mathcal{U}$
is Poisson distributed with mean $\lambda A t \P\{ V \in \mathcal{U} \}$.
The probability that the mutant type has not reached $x$ by time $t$ is
the exponential of the total measure of {\em possible} mutants that would have reached $(x,t)$.
This measure can be written as
\begin{align}
  h(t) = \int_0^t \int_0^\infty d \omega(d) r^{d-1} q(u,r) dr du ,
\end{align}
so that $h(t)$ is also the {\em mean} volume in space-time swept out by an expanding, unobstructed mutation over $t$ time units.
Just as before,
\begin{align}
  \P\{ \tau > t \} = \exp\left( - \lambda h(t) \right) .
\end{align}
Furthermore, following the same reasoning as for equation \eqref{eqn:generaldensity},
the mean density of types is given by
\begin{align}
    \nu = \int_0^\infty \lambda \P \{ \tau > t \} dt = \lambda \E[\tau] .
\end{align}

Since $h(t)$ is the mean volume swept out over $t$ time units,
the equations for $\tau$ and $\nu$ are the same as in the deterministic case, 
except the path of the wave has been replaced by the path of the {\em mean} wave, in this sense.
In other words, in computing the mean density of types, we may replace the path of the wave by its mean path.

\section{Wave speeds for some fat-tailed kernels}
\label{apx:fourierspeeds}

Here we apply the theory of Section \ref{ss:generalspeed} to
the {\em stable} and the {\em stretched exponential} families of dispersal kernels.
We do not dwell on the justification for expressions of wave expansion profiles of different dispersal kernels,
preferring instead to take this as input into the theory,
but for the interested (or suspicious) reader,
here we outline how the expressions are arrived at, 
which follows \citet{kot96}, with the minor addition of Equation \eqref{eqn:stable_tail}.
In each case, an expression for the wave speed is arrived at as follows.
Suppose that there is a population of size $N$ at every point,
and that mutant organisms have first a dispersal stage, where they disperse a distance according to a distribution $k(\cdot)$,
then undergo density-dependent growth, with a population of size $n(x)$ at location $x$ growing to $F(n(x))$.
The ensuing discrete-time integrodifference analogue to the Fisher--KPP equation studied e.g. in \cite{kot96} is
\[
    n(t+1,x) = \int k(x-y) F( n(t,y) ) dy .
\]
As is common in the study of such nonlinear equations,
we then linearize the equation by assuming that the spread of the wave only depends on the behavior when the mutant type is rare,
replacing $F(n)$ with $n F'(0)$, and writing $F'(0)=(1+s)$.  
See \cite{kot96} or \cite{mollison1972spatial} for discussion of this assumption.
Then the number of mutants $n$ satisfies
\[
    n(t+1,x) = (1+s) \int k(x-y) n(t,y) dy .
\]
Since the Fourier transform takes convolutions to products,
if we write 
$\widetilde n(t,\omega) = \int e^{i x \omega} n(t,x) dx$ 
and 
$\widetilde k(\omega) = \int e^{ix\omega} k(x) dx$
for the Fourier transforms of $n$ and $k$ respectively,
then
\begin{equation} \label{eqn:fourier_solution}
  \widetilde n(t,\omega) = (1+s)^t \widetilde n(0,\omega) \widetilde k(\omega)^t .
\end{equation}
One might hope to then obtain information about the spread of the wave from this expression,
or even by explicitly inverting it.
We will assume that the speed in two dimensions is the same as in one dimension, 
which can be easily proved at least in some sense for many of these models.

\subsection{Stretched exponential}
\label{apx:stretched}

A family of dispersal kernels with moments but without moment generating functions is
the family sometimes called ``stretched exponentials'', whose probability density function is
\[  k(x) = \frac{ \alpha c_\alpha }{ 2 \Gamma(1/\alpha) } e^{- |c_\alpha x|^\alpha } , \]
for $0<\alpha<1$.
If $\alpha=1$ and $d=1$ this is the Laplace distribution; more generally the distribution is sometimes also called the ``error distribution''.
Here $\alpha$ controls the decay of the tails
and $c_\alpha$ is a positive constant depending on $\alpha$ chosen so that the IQR matches that of the standard Gaussian.
This distribution with $\alpha=\frac12$ has been used by \citet{Novembre2005} for human dispersal, 
and \cite{kot96} showed that it gave the best fit out of a few choices to dispersal data for {\it Drosophila pseudoobscura}.
We introduce the scaling parameter $\sigma$ by replacing $k(x)$ with $k(x/\sigma)/\sigma$, as usual.

Following the reasoning above, \cite{kot96} argued that
the stretched exponential dispersal gives a wave that accelerates with a power of $t$:
\begin{equation}
f(t) \approx \sigma ( t s )^{1/\alpha} / c_\alpha .
\end{equation}
Because of this form, properties of the constant-speed case can be derived from the case $\alpha=1$, after substituting $v$ for $\sigma s / c_\alpha$
The characteristic length for this family of distributions has the following form:
\begin{equation}
\chi = \left( \frac{ s \sigma^\alpha }{ \lambda C } \right)^{1/(d+\alpha)} = \left( \frac{ \xi^2 \sigma^\alpha }{ 2 \rho \mu C } \right)^{1/(d+\alpha)} ,
\end{equation}
where $C = \omega(d) c_\alpha^{1/\alpha}$ is a constant.
Note that this is independent of $s$, unlike the constant-speed case.

In the remainder of this section we neglect the demographic parameters $r$ and $\xi^2$,
assuming for instance that the offspring distribution is Poisson with mean and variance equal to 1.
It is fairly straightforward to factor them in, but for simplicity we omit them.
The variable $r$ may appear, used as a radius; it should not cause confusion.

Using the fact that
\[
\int_0^\infty t^a e^{-b t^c} dt = \frac{b^{-(a+1)/c}}{c} \Gamma\left( \frac{a+1}{c} \right) ,
\]
where $\Gamma$ is the gamma function, we can compute that
\begin{gather*}
  h(t) = 
            \frac{ \alpha \omega(d)^{1+d/\alpha} }{ d+\alpha } (\chi^d t)^{(d+\alpha)/\alpha} , \\
  f^{-1}(r) = 
            \left( \frac{ r }{ \chi^d } \right)^\alpha , \\
  h \left( f^{-1}(r) \right) = \frac{ \alpha \omega(d)^{d/(d+\alpha)} }{ d+\alpha } (r/\chi)^{d+\alpha} ,
\end{gather*}
and apply the results of Section \ref{ss:generalspeed}.
Therefore, the mean area occupied by a typical mutation is
\begin{align*}
    \E[A] 
          &= \left( \int_0^\infty \exp(-h(t)) dt \right )^{-1} \\
          &= \chi^{d} \;  \omega(d) \left( \frac{\alpha}{d+\alpha} \right)^{-d/(d+\alpha)}
                \Gamma\left(\frac{\alpha}{d+\alpha}\right)^{-1}   ,
\end{align*}
Recall also that the mean time until a sample point is occupied satisfies
\[
    \E[\tau] = \frac1{\lambda \E[A]},
\]
which gives an idea of the time scale the process happens on.
The factor of $\lambda$ appears because of our time scaling.

Also, the moments of the distance $X$ of a sample point to the origin of its mutation have a nice form:
\begin{align} \label{eqn:explmoments}
  \begin{split}
      \E[X^n] &=  \frac{ d \omega(d) }{ d+n } \int_0^\infty r^{d+n} e^{-h(r)} dr \\
      &= \left( \frac{ s \sigma^\alpha d }{ \lambda \alpha } \right)^{\frac{d+n+1}{d+\alpha}}
          \frac{ d \omega(d) }{ (d+n)(d+\alpha) }
          \Gamma\left(\frac{d+n+1}{d+\alpha}\right) \\
        &= \chi^{d+n+1} \left( \frac{ d }{ \alpha } \right)^{\frac{d+n+1}{d+\alpha}}
          \frac{ d \omega(d) }{ (d+n)(d+\alpha) }
          \Gamma\left(\frac{d+n+1}{d+\alpha}\right) .
  \end{split}
\end{align}

\subsection{Stable distributions}
\label{apx:stable}

For another example of fat-tailed dispersal distributions, 
we take the well-known family of symmetric stable distributions,
which are parameterized by a scaling exponent $0 < \alpha < 2$.
Stable distributions arise naturally in the generalization of the central limit theorem 
and as scaling limits for random walks with step distributions
whose tails have power-law decay like $x^\alpha$ (so-called ``L\'evy walks''), and so are a natural choice for a dispersal distribution.
For recent discussion of modeling real dispersal with such distributions, see \citet{reynolds2008levywalk} and \citet{edwards2007revisiting}. 
The best-known example is the Cauchy distribution ($\alpha=1$).
The case $\alpha=2$ corresponds in some senses to the Gaussian distribution,
but what follows does not apply in that case because Equation \eqref{eqn:stable_tail} for the tail behavior does not hold.

Denote by $k_\alpha(x)$ the density function of the $\alpha$-stable distribution,
again normalized to have the same IQR as the standard Gaussian.
To incorporate a scaling parameter analogous to the standard deviation,
we use the dispersal kernel $k_\alpha(x/\sigma)/\sigma$.

In the case of a stable dispersal distribution, the Fourier transform \eqref{eqn:fourier_solution} can be explicitly inverted.
If we begin with $n_0$ mutants at the origin (as a delta distribution), then
the spread is itself given by the dispersal kernel,
\[
   n(t,x) = n_0 \frac{ (1+s)^t }{\sigma t^{1/\alpha}} k_\alpha\left( \frac{x}{\sigma t^{1/\alpha}} \right),
\]
so if $x_t>0$ is such that $n_t(x_t)=\epsilon n_0$, then
\[
    x_t = \sigma t^{1/\alpha} k_{\alpha}^{-1}\left( \frac{ \epsilon \sigma t^{1/\alpha} }{ (1+s)^t } \right) .
\]
The density of the stable distribution for general $\alpha$ is not known in general,
but its tail behaviour is.  For large $x$,
\begin{equation} \label{eqn:stable_tail}
   k_\alpha(x) \approx \frac{ \alpha \sin(\pi\alpha/2) \Gamma(\alpha) }{ \pi |x|^{1+\alpha} },
 \end{equation}
and so for large $t$ and small $\epsilon$,
\[
    x_t \approx \sigma \left( \frac1{ \epsilon \pi } t (1+s)^t \alpha \sin(\pi \alpha/2) \Gamma(\alpha) \right)^{1/(1+\alpha)} ,
\]
giving $f(t) = \sigma \left( t (1+s)^t \alpha \sin(\pi \alpha/2) \Gamma(\alpha) /\epsilon \right)^{1/(1+\alpha)}$.
Note that the asymptotic speed is no longer independent of $\epsilon$ --
positions farther out the wave front move faster.
For our numerical purposes, we will take $\epsilon=0.2$.

An analytic expression for the characteristic length can be written using the Lambert W--function,
but it is more efficient to solve \eqref{eqn:chidefn} numerically.
The integrals giving other properties of the process are difficult to do explicitly,
but can be done numerically as well.


\bibliography{references}

\end{document}